\documentclass[12pt]{article}

\usepackage{epsfig}
\usepackage{amssymb}
\usepackage{graphicx}
\usepackage{color}
\usepackage{subfigure}
\usepackage{mathtools}
\usepackage{latexsym}
\usepackage[hidelinks]{hyperref}

\makeatletter
\renewcommand{\theequation}{\thesection.\arabic{equation}}
\@addtoreset{equation}{section} \makeatother
\renewcommand{\vec}[1]{\underline{#1}}

\def\a{\alpha}

\def\d{\delta}

\def\ph{\phi}
\def\Ph{\Phi}

\def\m{\mu}

\def\s{\sigma}
\def\S{\Sigma}

\def\O{\Omega}

\def\lt{\left}
\def\rt{\right}

\def\nn{\nonumber}

\def\p{\partial}

\def\la{\langle}
\def\ra{\rangle}

\def\nn{\nonumber}

\def\bea{\begin{eqnarray}}
\def\eea{\end{eqnarray}}

\begin{document}

\begin{titlepage}
\title{\vskip -60pt
\vskip 20pt Vacua and walls of mass-deformed K\"{a}hler nonlinear sigma models on $SO(2N)/U(N)$
}
\author{
Bum-Hoon Lee$^a$\footnote{e-mail : bhl@sogang.ac.kr}, Chanyong
Park$^{bc}$\footnote{e-mail : chanyong.park@apctp.org }, Sunyoung
Shin$^{b}$\footnote{e-mail : sunyoung.shin@apctp.org}}
\date{}
\maketitle \vspace{-1.0cm}
\begin{center}
~~~
\it $^a\,$ Department of Physics, Sogang University, Seoul 121-742, Korea\\
\it $^b\,$Asia Pacific Center for Theoretical Physics, Pohang 790-784, Korea\\
\it $^c\,$ Department of Physics, Pohang University of Science and Technology,\linebreak Pohang, 790-784, Korea
\end{center}

\thispagestyle{empty}

\begin{abstract}
We construct walls of mass-deformed K\"{a}hler nonlinear sigma models on $SO(2N)/U(N)$, by using the moduli matrix formalism and the simple roots of $SO(2N)$. Penetrable walls are observed in the nonlinear sigma models on $SO(2N)/U(N)$ with $N>3$.
\end{abstract}

\end{titlepage}

\newpage


\section{Introduction} \label{sec:intro}
\setcounter{equation}{0}

The moduli matrix formalism which was developed in \cite{Isozumi:2004jc,Isozumi:2004va}, is one of powerful tools to study
Bogomol'nyi--Prasad--Sommerfield (BPS) objects. The moduli matrices contain all the moduli parameters of BPS solutions. In \cite{Isozumi:2004jc,Isozumi:2004va}, moduli matrices of walls are constructed in a mass-deformed hyper-K\"{a}hler nonlinear sigma model whose target space is the cotangent bundle over the complex Grassmann manifold $T^\ast G_{N_F,N_C}$, which is the strong
coupling limit of supersymmetric $U(N_C)$ gauge theory with $N_F>N_C$\footnote{This
condition is required to define the Grassmann manifold $G_{N_F,N_C}=\frac{SU(N_F)}{SU(N_C)\times
SU(N_F-N_C)\times U(1)}$.}  \cite{Arai:2003tc}. It is also shown that the vacua and the walls are on the
K\"{a}hler manifold. 

In the moduli matrix formalism, walls are generated by operators, which are positive root generators of the system. Elementary wall operators are generators of simple roots \cite{Sakai:2005sp}. Thus elementary walls and compressed walls can be labelled by simple roots and by linear combinations of simple roots respectively. We can also figure out structures of multiwalls from the linear combinations of simple roots.

Various BPS obejcts are constructed in moduli matrix formalism. Domain walls and domain wall webs are discussed in \cite{Eto:2004vy,Arai:2011gg,Eto:2005cp}. Other BPS objects and composite objects are discussed in \cite{Eto:2005yh,Eto:2009bg,Isozumi:2004vg,Eto:2004rz,Eto:2008mf,Arai:2016sdz,Eto:2011cv}. 

Moduli matrices, which are constrained by a single quadratic constraint $SO(2N)$ or $Sp(N)$, are constructed for walls in \cite{Arai:2011gg} and for kink monopoles \cite{Eto:2011cv}. In \cite{Arai:2011gg}, walls of mass-deformed K\"{a}hler nonlinear sigma models on $SO(2N)/U(N)$ with $N=2,3$ are discussed. It is known that there are isomorphisms 
\bea%
SO(4)/U(2)\simeq CP^1, ~~SO(6)/U(3)\simeq CP^3.
\eea%
Therefore the nonlinear sigma models discussed in \cite{Arai:2011gg} are Abelian theories. In Abelian theories, the ordering of walls is absolute. In non-Abelian theories, however, walls can pass through each other. Penetrable walls are observed in nonlinear sigma models on the Grassmann manifold \cite{Isozumi:2004va}. 

The purpose of this paper is to construct walls of mass-deformed K\"{a}hler nonlinear sigma models on $SO(2N)/U(N)$ with $N>3$, which are non-Abelian theories. We use the convention and part of the formalism used in \cite{Eto:2011cv}. In Section \ref{sec:sec2}, we review the moduli matrix method discussed in \cite{Isozumi:2004va,Arai:2011gg,Eto:2011cv}. In 
Section \ref{sec:n=3}, we review the results of \cite{Arai:2011gg} with the $O(2N)$ invariant tensor, which is used in \cite{Eto:2011cv}. We also identify elementary walls with the simple roots of $SO(2N)$. In Section \ref{sec:n=4}, we study elementary walls of nonlinear sigma models on $SO(2N)/U(N)$ with $N=4,5,6,7$, which are labelled by the simple roots of $SO(2N)$. In Section \ref{sec:Ngeneral}, we discuss the vacuum structure connected to the maximum number of elementary walls. In Section \ref{sec:observation}, we make some observations about walls of nonlinear sigma model on $SO(12)/U(6)$, which is the simplest nontrivial case. In Section \ref{sec:discuss}, we summarize our result.

%
\section{Moduli matrices on $SO(2N)/U(N)$}\label{sec:sec2}
\setcounter{equation}{0}
In this section, we review the formalism of \cite{Arai:2011gg,Eto:2011cv}. The mass-deformed K\"{a}hler nonlinear sigma model on $SO(2N)/U(N)$ in (2+1) dimensions \cite{Arai:2011gg} is obtained by dimensional reduction \cite{Scherk:1978ta} of $\mathcal{N}=1$ (3+1)-dimensional K\"{a}hler nonlinear sigma model on $SO(2N)/U(N)$ \cite{Higashijima:1999ki}. As we are interested in solitonic objects, we consider only the bosonic part of the model. The Lagrangian in (2+1) dimensions is
\bea
{\mathcal{L}}&=&-|D_\m\ph_a^{~i}|^2-|i\ph_a^{~j}M_j^{~i}-i\S_a^{~b}\ph_b^{~i}|^2
+|F_a^{~i}|^2+(D_a^{~b}\ph_b^{~i}\bar{\ph}_i^{~a}-D_a^{~a})\nn\\
&&+\lt((F_0)^{ab}\ph_b^{~i}J_{ij}\ph^{Tj}_{~~a}+\ph_0^{ab}F_b^{~i}J_{ij}\ph^{Tj}_a
+(\ph_0)^{ab}\ph_b^{~i}J_{ij}F^{Tj}_{~~\,a}+\mathrm{(h.c)}  \rt), \label{eq:3dlag}
\eea
where the Greek letter $\mu$ denotes three-dimensional spacetime indices and the covariant derivative is defined by $(D_\m\ph)_a^{~i}=\p_\m\ph_a^{~i}-iA_{\m a}^{~~\,b}\ph_b^{~i}$. The indices $i,j(=1,\cdots, 2N)$ are flavour indices and the indices $a,b(=1,\cdots,N)$ are color indices. The last term (h.c) stands for the Hermitian conjugate. The $O(2N)$ invariant tensor $J$ is defined by
\begin{eqnarray}
J=\s^1\otimes I_N, \label{eq:so2n_inv_ten}
\end{eqnarray}
where $I_N$ stands for the $N\times N$ identity matrix. In this basis, the $SO(2N)$ Cartan generators are
\bea
H_{n}=\left(
\begin{array}{c|c}
h_n & ~ \\ \hline ~   & -h_n
\end{array}
\right),~~(n=1,\cdots,N),
\eea
with $N\times N$ matrix $h_n$ which has an only component 1 in $(n,n)$-element. The mass matrix is a linear combination of the Cartan generators. By defining vectors
\bea
&&\vec{m}=(m_1,m_2\cdots,m_N),\nn\\
&&\vec{H}=(H_1,H_2,\cdots,H_N), \label{eq:m&H}
\eea
the mass matrix is formulated as
\bea
M=\vec{m}\cdot\vec{H}. \label{eq:mass1}
\eea
In the basis of $SO(2N)$ with the invariant tensor (\ref{eq:so2n_inv_ten}), the mass matrix (\ref{eq:mass1}) is
\bea
M=\s_3\otimes\mathrm{diag}(m_1,m_2,\cdots,m_N). \label{eq:mass2}
\eea
The constraints of the Lagrangian (\ref{eq:3dlag}) are
\bea
&&\ph_a^{~i}\bar{\ph}_i^{~b}-\d_a^{~b}=0, \label{eq:const1}\\
&&\ph_a^{~\,i}J_{ij}\ph^{Tj}_{~~b}=0,~~
\mbox{(hermitian conjugate)}=0.
\label{eq:const2}
\eea
Eq.(\ref{eq:const1}) is from the D-term constraint which defines the Grassmann manifold $G_{2N,N}$ and eqs.(\ref{eq:const2}) are from the F-term constraints which correspond to the (anti-)holomorphic embedding of $SO(2N)$ manifold.
By eliminating auxiliary fields, the potential term of the Lagrangian becomes
\bea
V=|i\ph_a^{~j}M_j^{~i}-i\S_a^{~b}\ph_b^{~i}|^2+4|(\ph_0)^{ab}\ph_b^{~i}|^2.
\eea
Therefore the vacuum conditions are
\bea
&&\ph_a^{~j}M_j^{~i}-\S_a^{~b}\ph_b^{~i}=0, \label{eq:vac}\\
&&(\ph_0)^{ab}=0.
\eea
$\S$ can be diagonalized as
\bea
\S=\mathrm{diag}(\S_1,\S_2,\cdots,\S_N), \label{eq:sigma}
\eea
by a $U(N)$ gauge transformation. Then the vacua are labelled by
\bea
(\S_1,\S_2,\cdots,\S_N)=(\pm m_1,\pm m_2,\cdots,\pm m_N). \label{eq:vaclab}
\eea
Since the tensor (\ref{eq:so2n_inv_ten}) is invariant under $O(2N)$ transformation, only the half of the vacua in (\ref{eq:vaclab}) belong to a nonlinear sigma model and the other half belong to the other nonlinear sigma model, which is related by parity. Therefore the number of vacua of the model is $2^{N-1}$. This number is the Euler characteristic of the manifold on which the nonlinear sigma model lives \cite{Witten:1982df,Hori:2000kt,SCM,Gudnason:2010rm}.

The BPS equation for wall solutions is derived from the Bogomol'nyi completion of the Hamiltonian. It is assumed that fields are static and all the fields depend only on $x_1\equiv x$ coordinate. It is also assumed that there is Poincar\'{e} invariance on the two-dimensional worldvolume of walls to set $A_0=A_2=0$. The energy density along the $x$-direction is
\bea
{\mathcal{E}}&=& \lt(|(D\ph)_a^{~i}|^2+|\ph_a^{~j}M_j^{~i}-\S_a^{~b}\ph_b^{~i}|^2+4|(\ph_0)^{ab}\ph_b^{~i}|^2\rt)  \nn\\
&=& \lt(|(D\ph)_a^{~i}\mp(\ph_a^{~j}M_j^{~i}-\S_a^{~b}\ph_b^{~i})|^2
+4|(\ph_0)^{ab}\ph_b^{~i}|^2\rt)\pm T  \nn\\
&\geq& \pm T.
\eea
This is also constrained by (\ref{eq:const1}) and (\ref{eq:const2}). The tension of the wall is
\bea
T= \p(\ph_a^{~i}M_i^{~j}\bar{\ph}_j^{~a}).
\eea
We choose the upper sign for the BPS equation and the lower sign
for the anti-BPS equation. The BPS equation is
\bea
(D\ph)_a^{~i}-(\ph_a^{~j}M_j^{~i}-\S_a^{~b}\ph_b^{~i})=0. \label{eq:bps}
\eea
By introducing complex matrix functions $S_a^{~b}(x)$ and $f_a^{~i}(x)$, which are defined by
\bea
\S_a^{~b}-iA_a^{~b}=(S^{-1}\p S)_a^{~b},~~~\ph_a^{~i}=(S^{-1})_a^{~b}f_b^{~i},
\eea
the equation (\ref{eq:bps}) is solved by
\bea
\ph_a^{~i}=(S^{-1})_a^{~b}H_{0b}^{~~j}(e^{Mx})_j^{~i}. \label{eq:bps_sol}
\eea
The coefficient matrix $H_0$ is called moduli matrix. $\S$, $A$ and $\ph$ are invariant under transformations
\bea
S_a^{\prime~b}=V_a^{~c}S_c^{~b},~~H_{0a}^{\prime~i}=V_a^{~c}H_{0c}^{~i}, \label{eq:worldvol}
\eea
where $V\in GL(N,{\mathbf{C}})$. The V defines an equivalent class of $(S,H_0)$. This is called the worldvolume symmetry \cite{Isozumi:2004jc,Isozumi:2004va}. The constraints (\ref{eq:const1}) and (\ref{eq:const2}) correspond to
\bea
&&H_{0a}^{~~i}(e^{2Mx})_i^{~j}H_{0j}^{\dagger~b}=(S\bar{S})_a^{~b}\equiv \O_a^{~b},\label{eq:hconst1}\\
&&H_{0a}^{~~i}J_{ij}H^{Tj}_{~~\,b}=0, ~\mbox{(hermitian conjuate)}=0, \label{eq:hconst2}
\eea
respectively. The worldvolume symmetry of $H_0$ in (\ref{eq:worldvol}) defines the Grassmann manifold. The constraint (\ref{eq:hconst2}) is a holomorphic embedding of $SO(2N)$. Therefore moduli matrices $H_0$'s parametrize $SO(2N)/U(N)$.

There are quantities that we can consider 
\bea
&&\mathrm{tr} \S=\frac{1}{2}\p\ln\det\O, \nn\\
&&T=\frac{1}{2}\p^2\ln \det \O.
 \label{eq:inv_quant}
\eea

In \cite{Isozumi:2004va}, walls are algebraically constructed from elementary walls. In the nonlinear sigma model on the Grassmann manifold, two nearest vacua have the same color number and differ by one flavour number. The elementary wall interpolating two nearest vacua, vacuum $\la A \ra$ and vacuum $\la B \ra$,
which are in the flavour $I$ and $I+1$ in the same color is $H_{0\la A\leftarrow B \ra}=H_{0\la A \ra}e^{E_I(r)}$ where $E_I(r)\equiv e^r E_I,~ (r\in{ \mathbf{C}})$. The elementary wall operator $E_I$ is defined by
\bea
[cM,E_I]=c(m_I-m_{I+1})E_I=T_{\la I\leftarrow I+1\ra}E_I, 
\label{eq:eledef1}
\eea
where $c$ is a constant, $M$ is the mass matrix and $E_I$ is an $N_f\times N_f$ square matrix generating an elementary wall. $T_{\la I\leftarrow I+1\ra}$ is the tention of the wall. The $E_I$ has a nonzero component only in the $(I,I+1)$-th element. However in the nonlinear sigma model on $SO(2N)/U(N)$, the definitions of the nearest vacua and the elementary wall are not valid due to the quadratic constraint (\ref{eq:hconst2}). Moduli matrices, which are constrained by a single quadratic constraint $SO(2N)$ or $Sp(N)$ are studied in \cite{Arai:2011gg,Eto:2011cv}. We review the methods for walls of nonlinear sigma models on $SO(2N)/U(N)$.

Since the mass matrix is a linear combination of the Cartan generators as it is defined in (\ref{eq:mass2}), we can generalize (\ref{eq:eledef1}) as
\bea
c[M,E_i]=c(\vec{m}\cdot \vec{\a}_i)E_i=T_iE_i. \label{eq:eledef2}
\eea
$E_i$ are the elementary wall operators which are the positive root generators of the simple roots $\a_i$ of $SO(2N)$. The elementary wall $H_{0\la A\leftarrow B \ra}=H_{0\la A \ra}e^{E_i(r)}$ where $E_i(r)\equiv e^r E_i,~ (r\in{ \mathbf{C}})$ can be labelled by the simple root $\a_i$. The wall has tension $T_i$.
We can restrict ourselves to the case where $m_1>m_2>\cdots>m_N$. Then the vector $\vec{m}$ of (\ref{eq:m&H}) is a vector in the interior of the positive Weyl chamber,
\bea
\vec{m}\cdot \vec{\a}_i>0.
\eea
Elementary walls can be identified with simple roots \cite{Sakai:2005sp}. The corresponding positive root generators are elementary wall operators. The formula (\ref{eq:eledef2}) is valid in nonlinear sigma models on the Grassmann manifold as well. 

We should only consider positive roots which correspond to the BPS solutions. Any linear combinations of positive roots and negative roots break supersymmetries. These linear combinations are called non-BPS sectors.

The simple roots of $SO(2N)$ are
\bea
&&\vec{\a}_1=(1,-1,0,\cdots,0,0,0),\nn\\
&&\vec{\a}_2=(0,1,-1,\cdots,0,0,0),\nn\\
&&\cdots \nn\\
&&\vec{\a}_{N-1}=(0,0,0,\cdots,0,1,-1),\nn\\
&&\vec{\a}_N=(0,0,0,\cdots,0,1,1). 
\label{eq:simplerootsN}
\eea
These simple roots describe elementary walls of nonlinear sigma models on $SO(2N)/U(N)$. The corresponding elementary wall operators are
\bea
&&E_i=
\begin{array}{c}  ~~~~ \\  i \\ ~ \\ ~ \\ i+N+1 \\ ~ \end{array}
\overset{\displaystyle \begin{array}{cccccc} ~ &\hspace{5mm} i+1 & ~ & ~& 
i+N & ~ \end{array}}
{\lt(
\begin{array}{ccc|ccc}
~ & ~ & ~ & ~ & ~ & ~  \\
~ & 1 & ~ & ~ & ~ & ~  \\
~ & ~ & ~ & ~ & ~ & ~  \\ \hline
~ & ~ & ~ & ~ & ~ & ~  \\
~ & ~ & ~ & ~ & -1 & ~  \\
~ & ~ & ~ & ~ & ~ & ~  
\end{array}\rt)}, ~~~ (i=1,\cdots, N-1),  \nn \\~\nn\\
&&E_N=
\begin{array}{c}  ~ \\ N-1  \\ N \\ ~ \\ ~ \\ ~ \end{array}
\overset{\displaystyle \begin{array}{cccccc} ~ &~ & ~ & ~&  2N-1 & 2N \end{array}}
{\lt(
\begin{array}{ccc|ccc}
~ & ~ & ~ & ~ & ~ & ~  \\
~ & ~ & ~ & ~ &  & 1  \\
~ & ~ & ~ & ~ & -1 & ~  \\ \hline
~ & ~ & ~ & ~ & ~ & ~  \\
~ & ~ & ~ & ~ & ~ & ~  \\
~ & ~ & ~ & ~ & ~ & ~  
\end{array}\rt)}.
\eea
The elementary wall, which connects two nearest vacua $\la A \ra$ and $\la B \ra$ is defined by 
\bea
&&H_{0\la A\leftarrow B\ra}=H_{0\la A \ra }e^{E_a(r)},\nn\\
&&E_a(r)\equiv e^rE_a,~~(a=1,\cdots,N),
\eea
where $E_a$ is the elementary wall operator and $r$ is a complex parameter ranging $-\infty<\mathrm{Re}(r)<\infty$. Elementary walls can be compressed to a single compressed wall. A compressed wall of level $n$, which connects $\la A \ra$ and $\la A^\prime \ra$ is defined by
\bea
&&H_{0\la A\leftarrow A^\prime\ra} =H_{0\la a \ra }e^{[E_{a_1},[E_{a_2},[E_{a_3},[\cdots,[E_{a_n},E_{a_{n+1}}]]]\cdots]](r)},\nn\\
&&(a_i=1,\cdots,N). 
\eea
%
%
There can exist multiwalls which connect two vacua $\la A\ra$ and $\la  B \ra$
\bea
H_{0\la A\leftarrow B \ra}=H_{0\la A \ra}e^{E_{a_1}(r_1)}e^{E_{a_2}(r_2)}\cdots e^{E_{a_n}(r_n)},
\eea
where parameters $r_i~(i=1,2,\cdots)$ are complex parameters ranging $-\infty < \mathrm{Re}(r_i) < \infty$. Elementary walls can pass through each other if
\bea
[E_{a_i},E_{a_j}]=0. 
\eea
These walls are called penetrable walls.

\section{Review on $SO(2N)/U(N)$ with $N=2,\,3$}\label{sec:n=3}
\setcounter{equation}{0}
In this section, we review the moduli matrices of vacua and walls of the nonlinear sigma models on $SO(2N)/U(N)$ with $N=2,\,3$ \cite{Arai:2011gg} by using the formalism of \cite{Eto:2011cv} and the simple roots of $SO(2N)$ (\ref{eq:simplerootsN}), which are summarized in Section \ref{sec:sec2}. The moduli matrices of the vacua are discussed in \ref{sec:app1}. We label the moduli matrices of vacua in descending order as follows:
\bea
&&(\S_1,\S_2,\cdots,\S_{N-1},\S_N)=(m_1,m_2,\cdots,m_{N-1},m_N), \nn\\
&&(\S_1,\S_2,\cdots,\S_{N-1},\S_N)=(m_1,m_2,\cdots,-m_{N-1},-m_N), \nn\\
&&~~~~~~~\vdots \nn\\
&&(\S_1,\S_2,\cdots,\S_{N-1},\S_N)=(\pm m_1,-m_2,\cdots,-m_{N-1},-m_N),
\eea
where the sign $\pm$ is $+$ for odd $N$ and $-$ for even $N$. $H_{0\la A\ra}$ and $\la A\ra$ denote a vacuum. 
$H_{0\la A_1 \leftarrow A_2 \leftarrow \cdots \leftarrow A_p\ra}$ and $\la A_1 \leftarrow A_2 \leftarrow \cdots \leftarrow A_p\ra$ denote a wall which connects vacua $\la A_1 \ra \leftarrow \la A_2 \ra \leftarrow \cdots \leftarrow \la A_p\ra$.

The vacua of the nonlinear sigma model on $SO(4)/U(2)$ (\ref{eq:modulimvac_2}) are
\bea
&&H_{0\la 1 \ra}=\lt(
\begin{array}{cc|cc}
1   &  ~   &   0   &   ~   \\
~   &  1   &   ~   &   0 
\end{array}
\rt),~~(\S_1,\S_2)=(m_1,m_2), \nn\\
&&H_{0\la 2 \ra}=\lt(
\begin{array}{cc|cc}
0   &  ~   &   1   &   ~   \\
~   &  0   &   ~   &   1 
\end{array}
\rt),~~(\S_1,\S_2)=(-m_1,-m_2).
\eea
There is only one elementary wall operator
\bea
E=\lt(
\begin{array}{cc|cc}
0   &  ~  &  0   &  1  \\
~   &  0  &  -1  &  0  \\ \hline
0   &  ~  &   0  &  ~  \\
~   &  0  &   ~  &  0  \\
\end{array}
\rt).
\eea
The elementary wall is
\bea
H_{0\la 1\leftarrow 2 \ra}=H_{0\la 1 \ra}e^{E(r)}
=\lt(
\begin{array}{cc|cc}
1   &  ~   &    0      &   e^r   \\
~   &  1   &   -e^r    &   0 
\end{array}
\rt).
\eea
This is the only wall of the nonlinear sigma model on $SO(4)/U(2)$. 

\begin{figure}[ht!]
\begin{center}
$\begin{array}{ccc}
\includegraphics[width=4.5cm,clip]{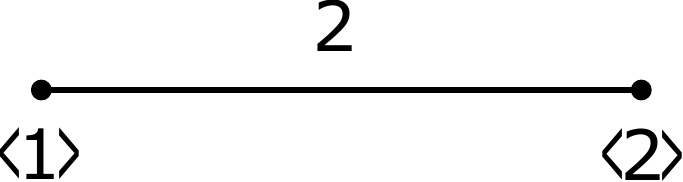}
&~~~~~~&
\includegraphics[width=7cm,clip]{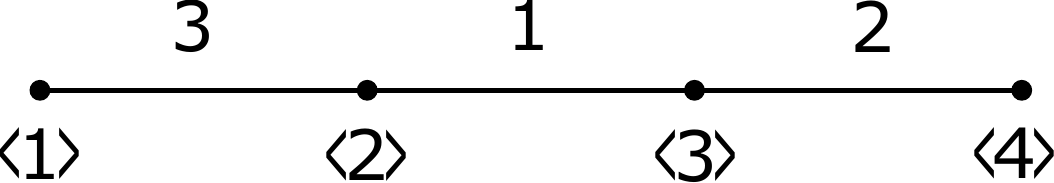}\\
\mathrm{(a)} &~~~~~~& \mathrm{(b)}
\end{array}
$
\end{center}
 \caption{(a)$SO(4)/U(2)$ (b)$SO(6)/U(3)$. The numbers indicate the subscript $i$'s of roots
  $\vec{\a}_i$. }
 \label{fig:n2n3rts}
\end{figure}

The vacua of the nonlinear sigma model on $SO(6)/U(3)$ (\ref{eq:modulimvac_3}) are 
\begin{eqnarray}
&&H_{0\la 1\ra}=\left(
\begin{array}{ccc|ccc}
1   &  ~  &  ~  &  0  & ~ & ~ \\
~   &  1  &  ~  &  ~  & 0 & ~ \\
~   &  ~  &  1  &  ~  & ~ & 0
\end{array}
\right),~~~(\S_1,\S_2,\S_3)=(m_1,m_2,m_3), \nn \\
&&H_{0\la 2\ra}=\left(
\begin{array}{ccc|ccc}
1   &  ~  &  ~  &  0  & ~ & ~ \\
~   &  0  &  ~  &  ~  & 1 & ~ \\
~   &  ~  &  0  &  ~  & ~ & 1
\end{array}
\right),~~~(\S_1,\S_2,\S_3)=(m_1,-m_2,-m_3),  \nn \\
&&H_{0\la 3\ra}=\left(
\begin{array}{ccc|ccc}
0   &  ~  &  ~  &  1  & ~ & ~ \\
~   &  1  &  ~  &  ~  & 0 & ~ \\
~   &  ~  &  0  &  ~  & ~ & 1
\end{array}
\right),~~~(\S_1,\S_2,\S_3)=(-m_1,m_2,-m_3), \nn \\
&&H_{0\la 4\ra}=\left(
\begin{array}{ccc|ccc}
0   &  ~  &  ~  &  1  & ~ & ~ \\
~   &  0  &  ~  &  ~  & 1 & ~ \\
~   &  ~  &  1  &  ~  & ~ & 0
\end{array}
\right),~~~(\S_1,\S_2,\S_3)=(-m_1,-m_2,m_3). 
\end{eqnarray}
The positive root generators of $SO(6)$, which are the elementary wall operators of $SO(6)/U(3)$ are
\bea
&&E_1=\left(
\begin{array}{ccc|ccc}
0   &  1  &  ~  &  ~  & ~ & ~ \\
~   &  0  &  ~  &  ~  & ~ & ~ \\
~   &  ~  &  0  &  ~  & ~ & ~ \\\hline
~   &  ~  &  ~  &  0  & ~ & ~ \\
~   &  ~  &  ~  &  -1 & 0 & ~ \\
~   &  ~  &  ~  &  ~  & ~ & 0
\end{array}
\right),~
E_2=\left(
\begin{array}{ccc|ccc}
0   &  ~  &  ~  &  ~  & ~ & ~ \\
~   &  0  &  1  &  ~  & ~ & ~ \\
~   &  ~  &  0  &  ~  & ~ & ~ \\\hline
~   &  ~  &  ~  &  0  & ~ & ~ \\
~   &  ~  &  ~  &  ~  & 0 & ~ \\
~   &  ~  &  ~  &  ~  & -1 & 0
\end{array}
\right),\nn\\
&&E_3=\left(
\begin{array}{ccc|ccc}
0   &  ~  &  ~  &  0  & ~ & ~ \\
~   &  0  &  ~  &  ~  & 0 & 1 \\
~   &  ~  &  0  &  ~  & -1 & 0 \\\hline
~   &  ~  &  ~  &  0  & ~ & ~ \\
~   &  ~  &  ~  &  ~  & 0 & ~ \\
~   &  ~  &  ~  &  ~  & ~ & 0
\end{array}
\right). \label{eq:rootgen3}
\eea
The elementary walls are 
\bea
&&H_{0\la 1 \leftarrow 2   \ra }=H_{0\la 1 \ra}e^{E_3(r_1)}=\left(
\begin{array}{ccc|ccc}
1   &  ~  &  ~  &  0  & ~ & ~ \\
~   &  1  &  ~  &  ~  & 0 & e^{r_1} \\
~   &  ~  &  1  &  ~  & -e^{r_1} & 0
\end{array}
\right), \nn\\
&&H_{0\la 2 \leftarrow 3   \ra }=H_{0\la 2 \ra}e^{E_1(r_1)}=\left(
\begin{array}{ccc|ccc}
1   &  e^{r_1}  &  ~  &  0     & ~   & ~ \\
~   &  0    &  ~  &  -e^{r_1}  & 1   & ~ \\
~   &  ~    &  0  &  ~     & ~   & 1
\end{array}
\right), \nn\\
&&H_{0\la 3 \leftarrow 4   \ra }=H_{0\la 3 \ra}e^{E_2(r_1)}=\left(
\begin{array}{ccc|ccc}
0   &  ~    &  ~    &  1    &  ~    & ~ \\
~   &  1    &  e^{r_1}  &  ~    &  0    & ~ \\
~   &  ~    &  0    &  ~    & -e^{r_1}  & 1
\end{array}
\right).
\eea
There are two compressed walls of level one. These are generated by following operators
\bea
\tilde{E}_4=[E_3,E_1],~\tilde{E}_5=[E_1,E_2].
\eea
We have used $\sim$ to distinguish the operators from elementary wall operators. The walls are
\bea
&&H_{0\la 1 \leftarrow 3 \ra}=H_{0\la 1 \ra}e^{\tilde{E}_4(r_1)}, \nn\\
&&H_{0\la 2 \leftarrow 4 \ra}=H_{0\la 2 \ra}e^{\tilde{E}_5(r_1)}.
\eea
There is only one compressed wall of level two, which is generated by
\bea
\tilde{E}_6=[E_3,\tilde{E}_5]=[\tilde{E}_4,E_2].
\eea
The moduli matrix for the wall is
\bea
H_{0\la 1 \leftarrow 4 \ra}=H_{0\la 1 \ra}e^{\tilde{E}_6(r_1)}.
\eea
There are four double walls
\bea
&&H_{0\la 1 \leftarrow 2 \leftarrow 3  \ra}
=H_{0\la 1 \leftarrow 2 \ra} e^{E_1(r_2)}
=\left(
\begin{array}{ccc|ccc}
1   &  e^{r_2}  &  ~     &   0    &    ~     &   ~   \\
~   &  1        &  ~     &   ~    &    0     &   e^{r_1} \\
~   &  ~        &  1     & e^{r_1+r_2} & -e^{r_1} &  0
\end{array}
\right), \nn\\
&&H_{0\la 2 \leftarrow 3 \leftarrow 4  \ra}
=H_{0 \la 2 \leftarrow 3 \ra}e^{E_2(r_2)}
=\left(
\begin{array}{ccc|ccc}
1   &  e^{r_1}  &  e^{r_1+r_2}     &   0    &    ~     &   ~   \\
~   &  0        &  ~               &   -e^{r_1}    &    1     &   ~ \\
~   &  ~        &  0               &   ~   & -e^{r_2} &  1
\end{array}
\right),
 \nn\\
&&H_{0\la 1 \leftarrow 2 \leftarrow 4  \ra}
=H_{0 \la 1 \leftarrow 2 \ra}e^{\tilde{E}_5(r_2)}
=\left(
\begin{array}{ccc|ccc}
1   &  ~        &  e^{r_2}     &   0            &    ~     &   ~   \\
~   &  1        &  ~           &   -e^{r_1+r_2} &    0     &   e^{r_1} \\
~   &  ~        &  1           &   ~            & -e^{r_1} &  0 
\end{array}
\right), \nn\\
&&H_{0\la 1 \leftarrow 3 \leftarrow 4  \ra}
=H_{0 \la 1 \leftarrow 3 \ra}e^{E_2(r_2)}
=\left(
\begin{array}{ccc|ccc}
1   &  ~        &  ~        &   0          &    e^{r_1+r_2} &  -e^{r_1}   \\
~   &  1        &  e^{r_2}  &   ~          &    0           &     ~ \\
~   &  ~        &  1        &   e^{r_1}    &    ~           &     0 
\end{array}
\right),
\eea
and one triple wall
\bea
H_{0 \la 1 \leftarrow 2 \leftarrow 3 \leftarrow 4 \ra}
=H_{0 \la 1 \leftarrow 2 \leftarrow 3 \ra}e^{E_2(r_3)}
=\left(
\begin{array}{ccc|ccc}
1   &  e^{r_2}  &  e^{r_2+r_3}   &   0         &   ~             &   ~       \\
~   &  1        &  e^{r_3}       &   ~         &   -e^{r_1+r_3}  &   e^{r_1} \\
~   &  ~        &  1             & e^{r_1+r_2} &   -e^{r_1}      &   0
\end{array}
\right).
\eea
We investigate the walls by using positive roots. The $SO(6)$ Cartan generators are
\bea
&&H_1=\mathrm{diag}(1,0,0,-1,0,0), \nn \\
&&H_2=\mathrm{diag}(0,1,0,0,-1,0),\nn \\
&&H_3=\mathrm{diag}(0,0,1,0,0,-1),
\eea
and the simple roots are
\bea
&&\vec{\a}_1:=(1,-1,0), \nn\\
&&\vec{\a}_2:=(0,1,-1), \nn\\
&&\vec{\a}_3:=(0,1,1). \label{eq:smplroots_3}
\eea
The subscripts correspond to the subscripts of the elementary wall operators in (\ref{eq:rootgen3}). 
We indicate the root which connects two vacua $\la i\ra \leftarrow \la j \ra$ as $g_{\la i \leftarrow j \ra}$. Then the elementary walls are
\bea
&&g_{\la 1 \leftarrow 2 \ra}=\vec{\a}_3, \nn\\
&&g_{\la 2 \leftarrow 3 \ra}=\vec{\a}_1,\nn\\
&&g_{\la 3 \leftarrow 4 \ra}=\vec{\a}_2,
\eea
and the roots of compressed walls are
\bea
&&g_{\la 1 \leftarrow 3 \ra}=\vec{\a}_3+\vec{\a}_1,  \nn\\
&&g_{\la 2 \leftarrow 4 \ra}=\vec{\a}_1+\vec{\a}_2,  \nn\\
&&g_{\la 1 \leftarrow 4  \ra}=\vec{\a}_1+\vec{\a}_2+\vec{\a}_3.
\eea
Diagrams of simple roots which connect the vacua of nonlinear sigma models on $SO(4)/U(2)$ and $SO(6)/U(3)$ are drawn in Fig.\ref{fig:n2n3rts}.
\vspace{1cm}

Once we identify the moduli matrices of vacua and the simple roots which connect the vacua, we can construct all the operators and the moduli matrices of walls. From the next section we concentrate only on the moduli matrices of vacua and the simple roots.

\section{$SO(2N)/U(N)$ with $N=4,5,6,7$}\label{sec:n=4} 
\setcounter{equation}{0}
As discussed in Section \ref{sec:intro}, nonlinear sigma models on $SO(2N)/U(N)$ with $N>3$ are non-Abelian theories. Therefore we expect that some of walls are penetrable. 

\begin{figure}[ht!]
\begin{center}
$\begin{array}{cc}
\includegraphics[width=7cm,clip]{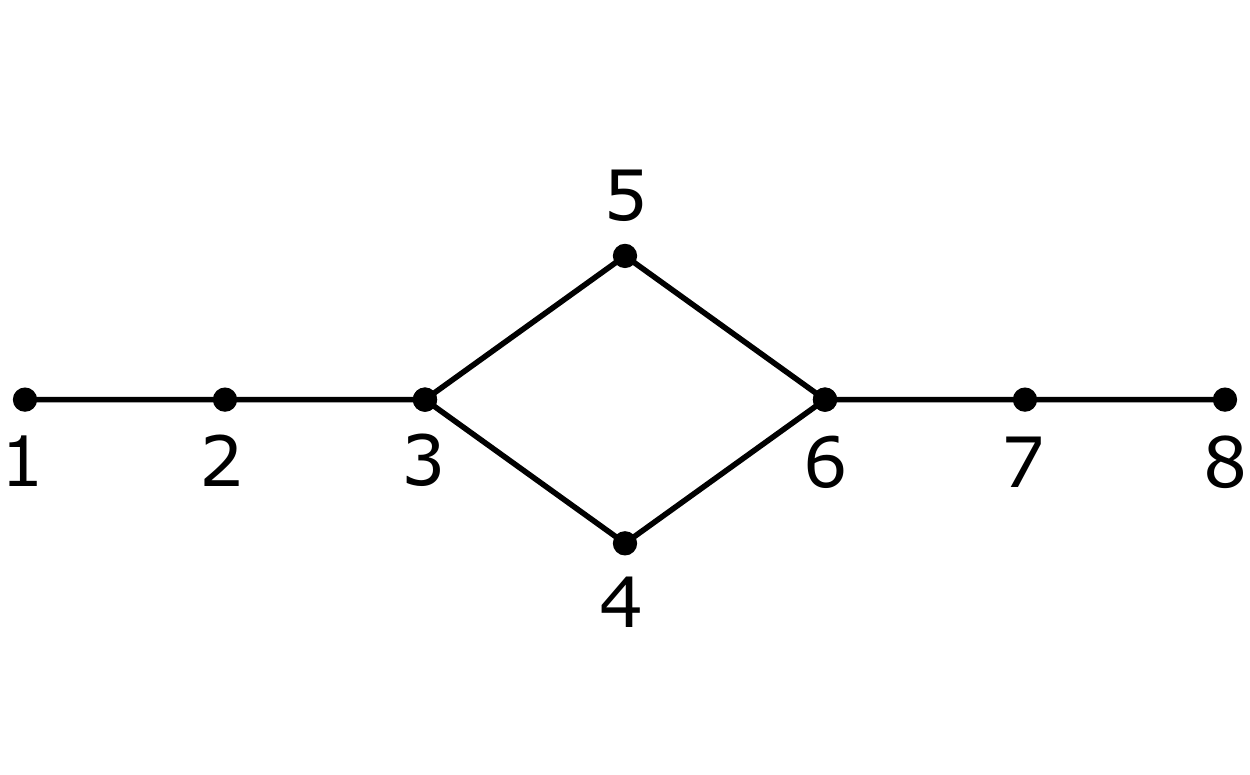}
&
\includegraphics[width=7cm,clip]{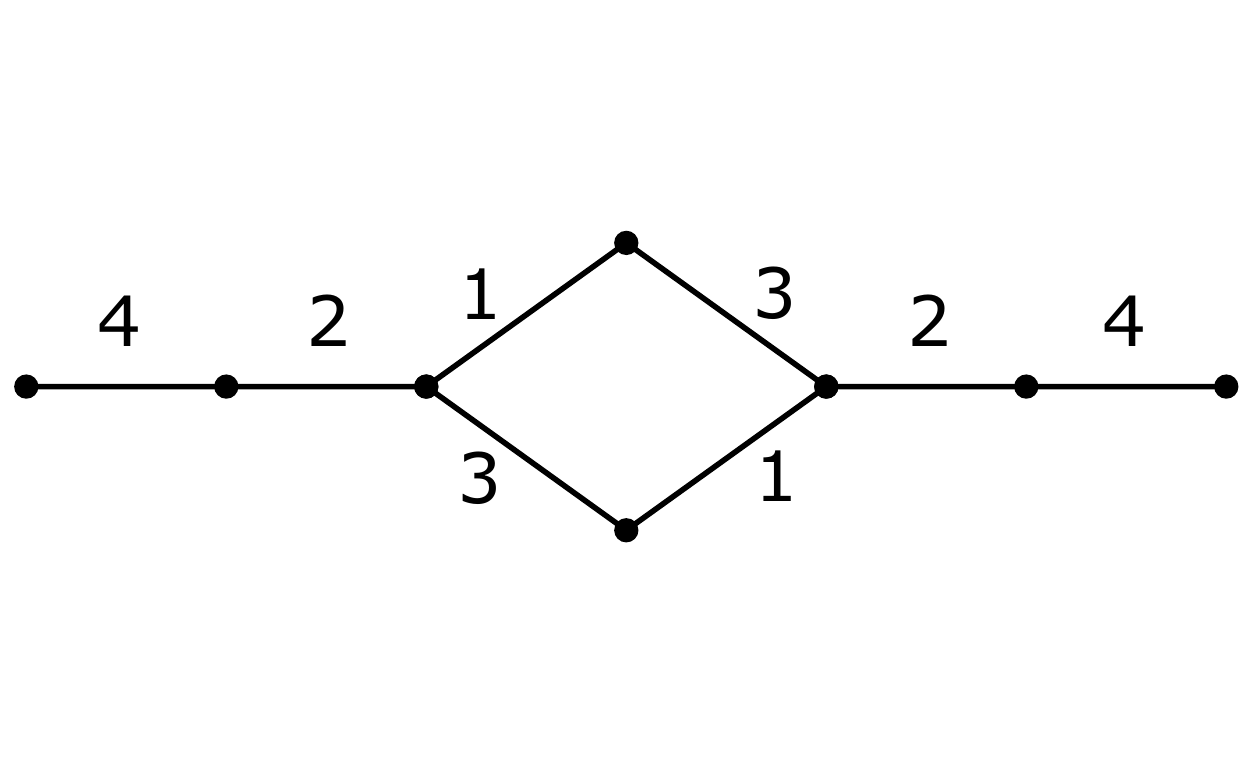}\\
\mathrm{(a)} & \mathrm{(b)}
\end{array}
$
\end{center}
 \caption{(a)Vacua of $SO(8)/U(4)$. The numbers indicate the labels of vacua. (b)Elementary walls of $SO(8)/U(4)$. The numbers indicate the subscript $i$'s of simple roots $\vec{\a}_i$, $(i=1,\cdots,4)$.
  }
 \label{fig:n4}
\end{figure}

\begin{figure}[ht!]
\vspace{2cm}
\begin{center}
$\begin{array}{ccc}
\includegraphics[width=5cm,clip]{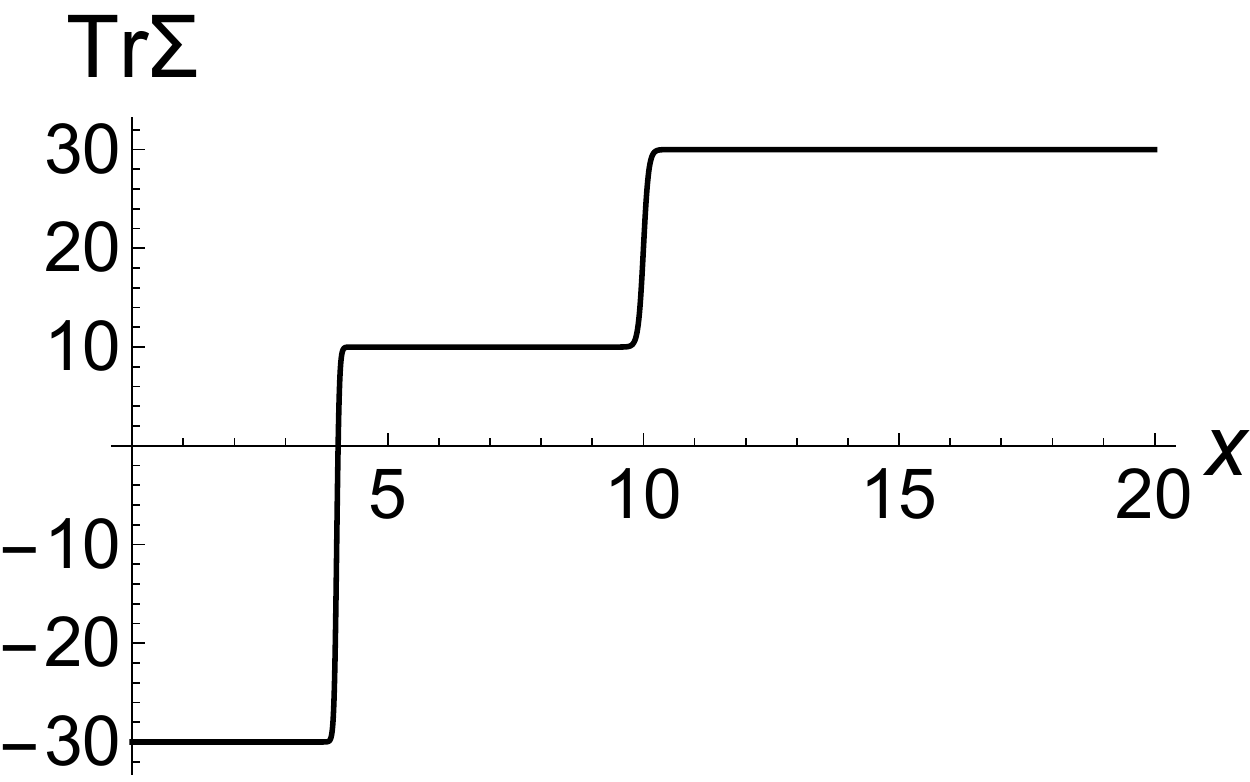}
&
\includegraphics[width=5cm,clip]{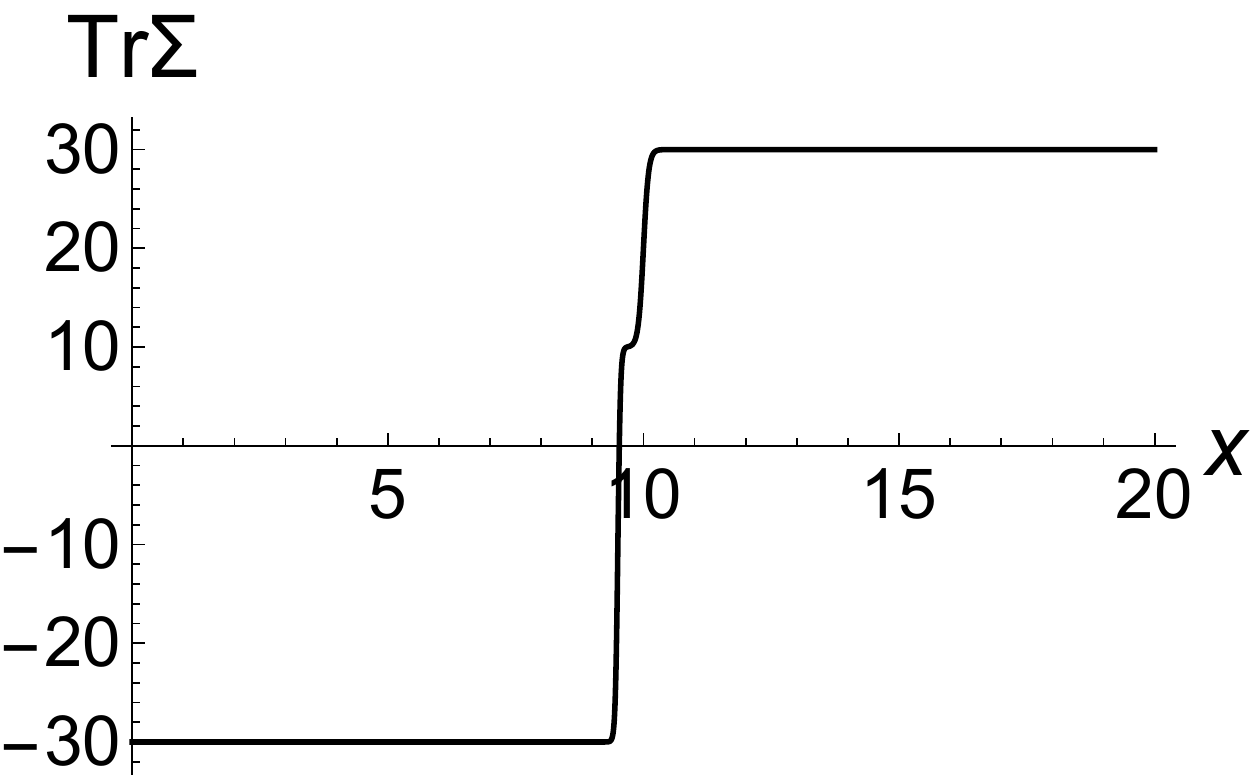}
& 
\includegraphics[width=5cm,clip]{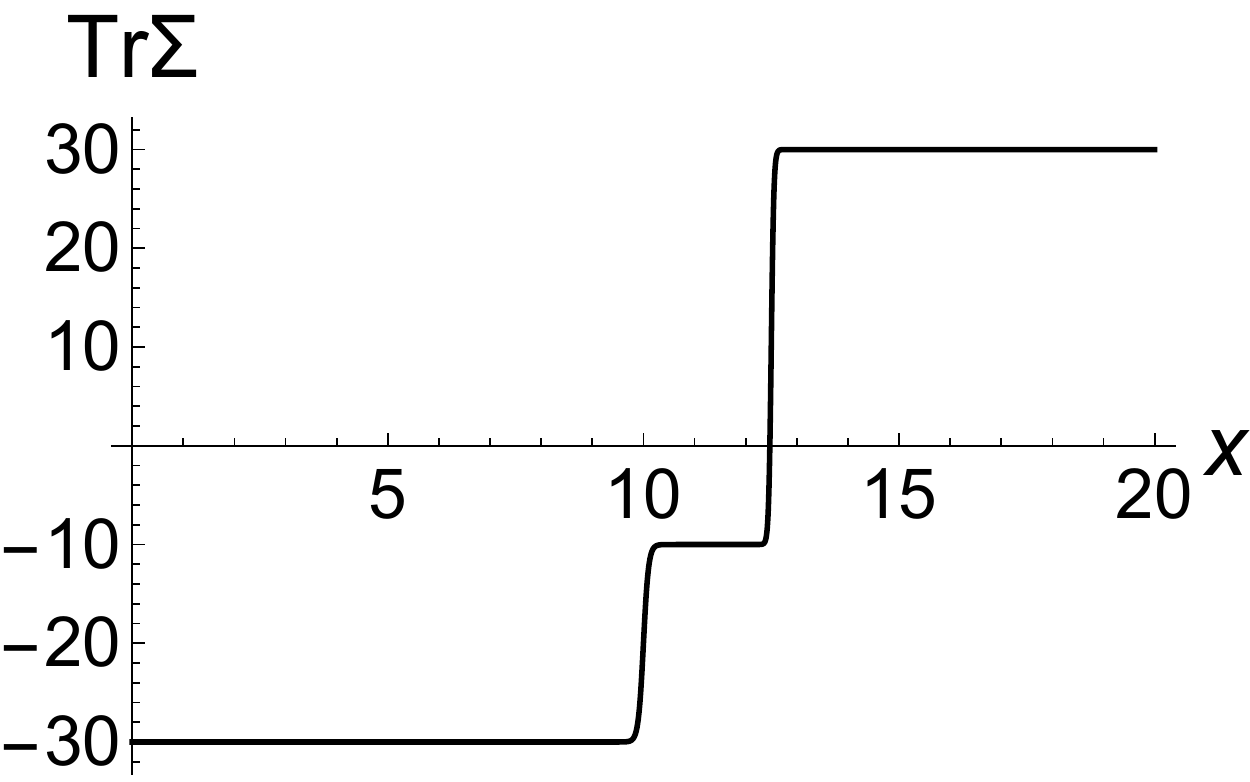} \\~~\\
\includegraphics[width=5cm,clip]{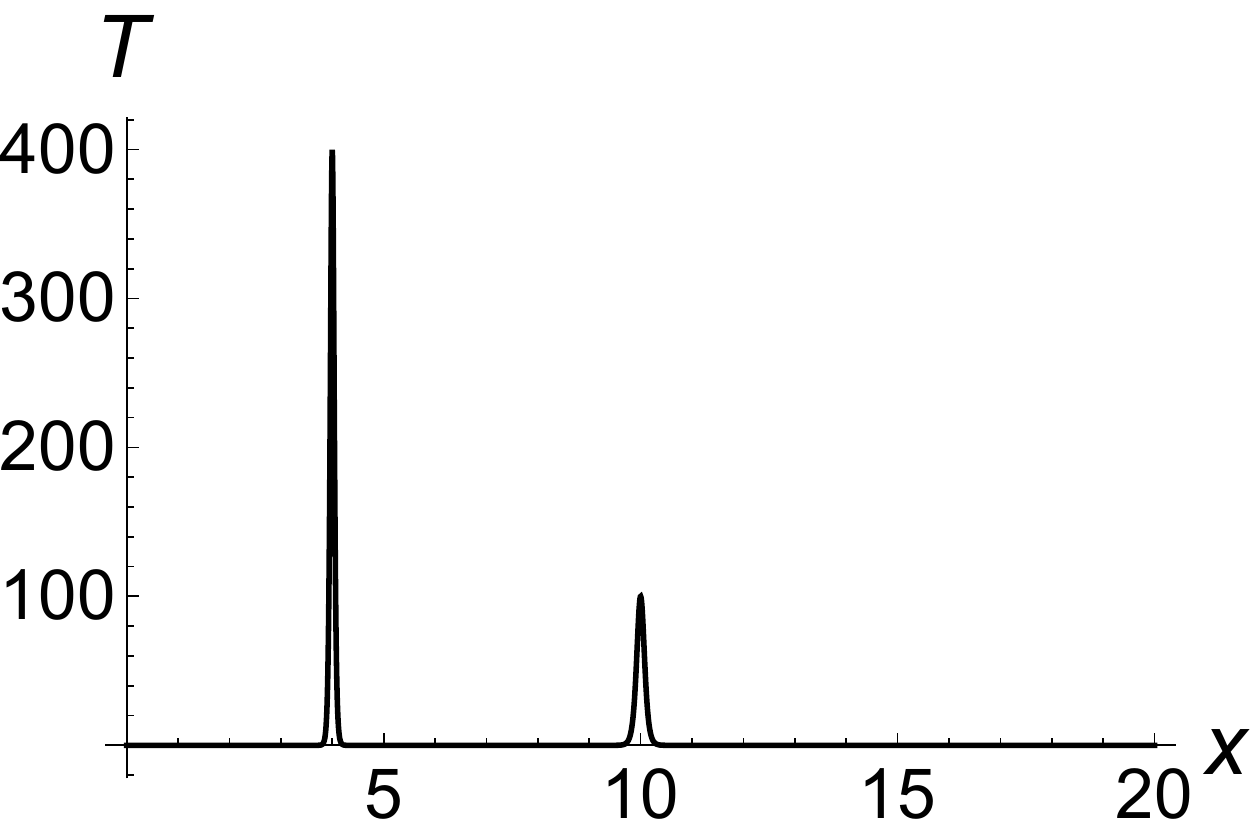}
&
\includegraphics[width=5cm,clip]{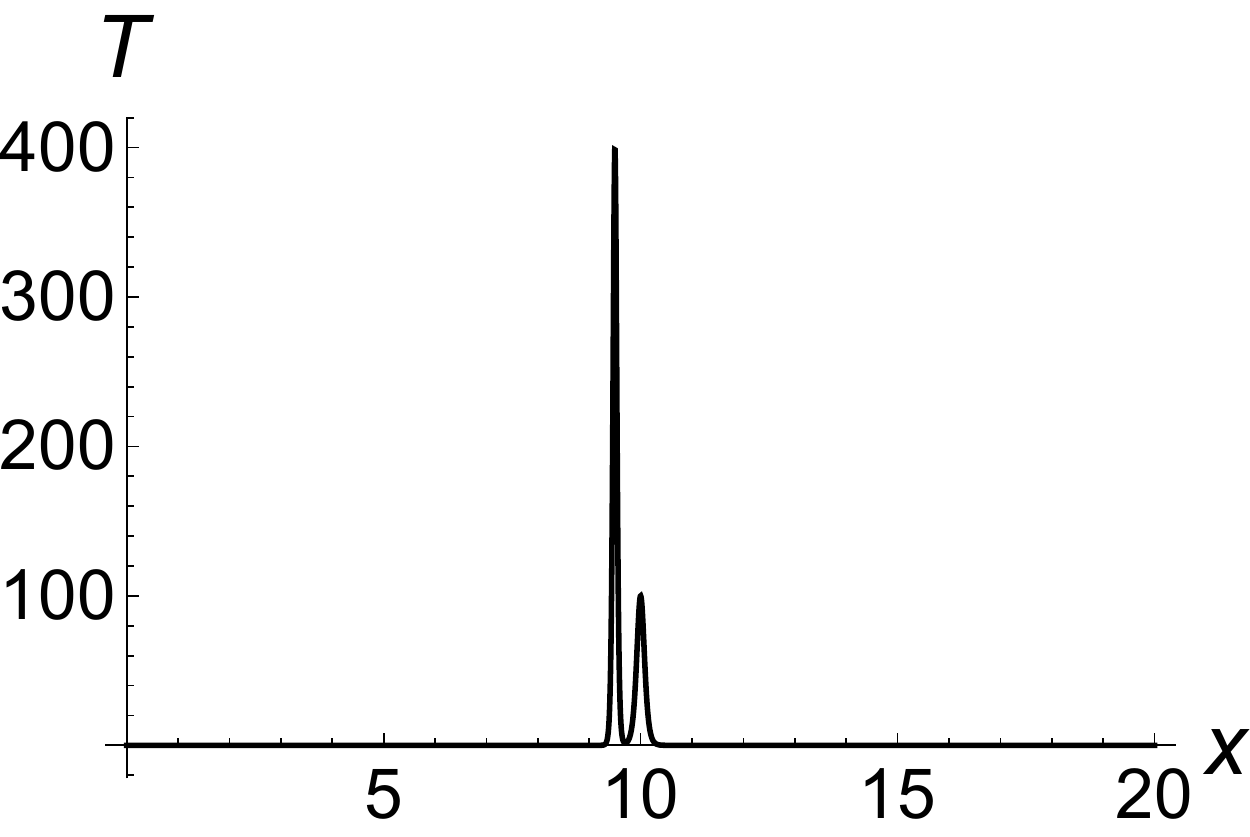}
& 
\includegraphics[width=5cm,clip]{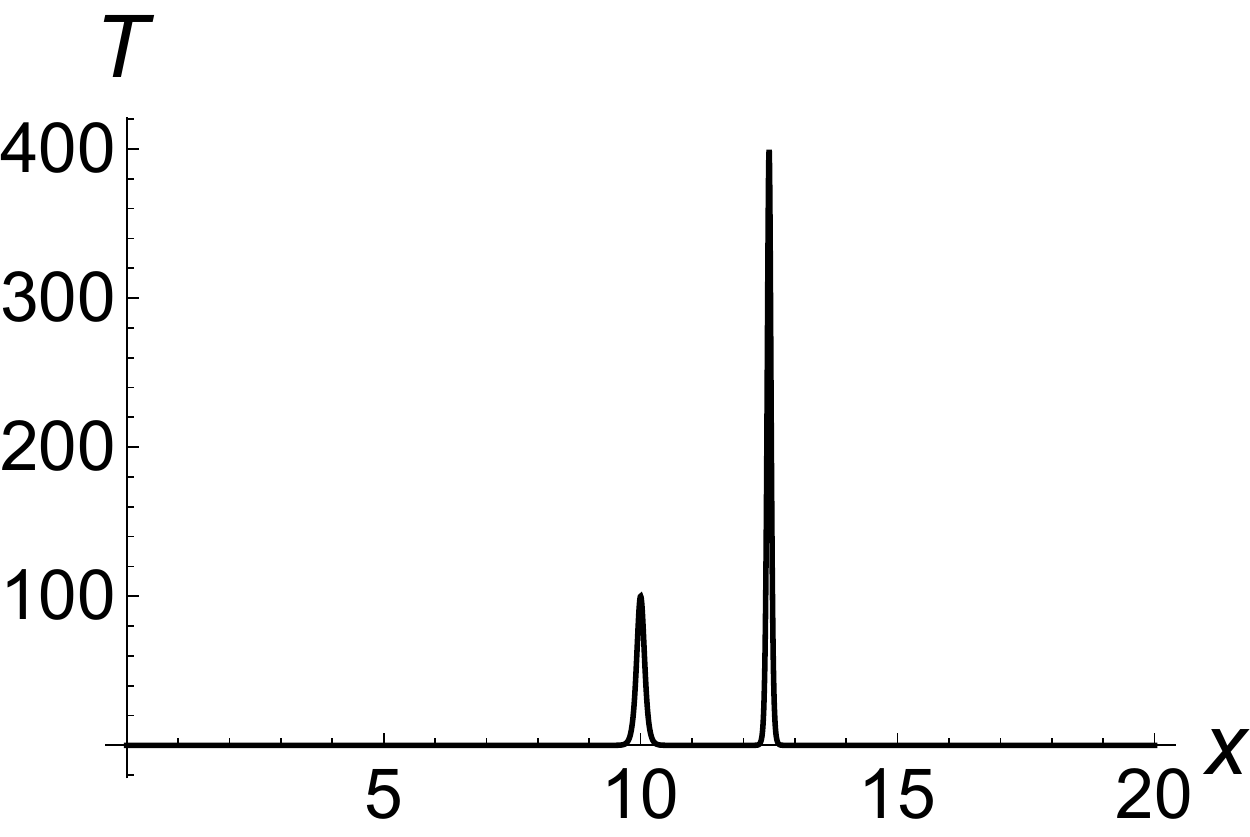}\\
\mathrm{(a)} & \mathrm{(b)} & \mathrm{(c)}
\end{array}
$
\end{center}
 \caption{Double wall $\la 3 \leftarrow  5  \leftarrow 6 \ra$ in $SO(8)/U(4)$, which consists of two penetrable walls. 
 $m_1=70$, $m_2=50$, $m_3=30$, $m_4=20$. (a)$r_1=80$, $r_2=100$ (b)$r_1=200$, $r_2=100$ (c)$r_1=250$, $r_2=100$.}
 \label{eq:n4h356}
\end{figure}

For $N=4$ there are $8(=2^3)$ vacua. The moduli matrices of vacua are defined by
\bea
&&H_{0\la 1 \ra}~:~
(\S_1,\S_2,\S_3,\S_4)=(m_1,m_2,m_3,m_4), \nn\\
&&H_{0\la 2 \ra}~:~
(\S_1,\S_2,\S_3,\S_4)=(m_1,m_2,-m_3,-m_4), \nn\\
&&H_{0\la 3 \ra}~:~
(\S_1,\S_2,\S_3,\S_4)=(m_1,-m_2,m_3,-m_4), \nn\\
&&H_{0\la 4 \ra}~:~
(\S_1,\S_2,\S_3,\S_4)=(m_1,-m_2,-m_3,m_4), \nn\\
&&H_{0\la 5 \ra}~:~
(\S_1,\S_2,\S_3,\S_4)=(-m_1,m_2,m_3,-m_4), \nn\\
&&H_{0\la 6 \ra}~:~
(\S_1,\S_2,\S_3,\S_4)=(-m_1,m_2,-m_3,m_4), \nn\\
&&H_{0\la 7 \ra}~:~
(\S_1,\S_2,\S_3,\S_4)=(-m_1,-m_2,m_3,m_4), \nn\\
&&H_{0\la 8 \ra}~:~
(\S_1,\S_2,\S_3,\S_4)=(-m_1,-m_2,-m_3,-m_4).
\eea
The simple roots of $SO(8)$ are
\bea
&&\vec{\a}_1:=(1,-1,0,0), \nn\\
&&\vec{\a}_2:=(0,1,-1,0), \nn\\
&&\vec{\a}_3:=(0,0,1,-1), \nn\\
&&\vec{\a}_4:=(0,0,1,1).
\eea
Then the roots of elementary walls $g_{\la i\leftarrow j  \ra}$ are
\bea
g_{\la 3\leftarrow 4  \ra}=g_{\la 5\leftarrow 6  \ra}=\vec{\a}_1, \nn\\
g_{\la 2\leftarrow 3  \ra}=g_{\la 6\leftarrow 7  \ra}=\vec{\a}_2, \nn\\
g_{\la 3\leftarrow 5  \ra}=g_{\la 4\leftarrow 6  \ra}=\vec{\a}_3, \nn\\
g_{\la 1\leftarrow 2  \ra}=g_{\la 7\leftarrow 8  \ra}=\vec{\a}_4.
\label{eq:so8rt}
\eea
The roots of penetrable walls are orthogonal
\bea
\vec{\a}_i\cdot \vec{\a}_j=0.
\eea
The vacua and the roots are depicted in Figure \ref{fig:n4}. A pair of orthogonal simple roots in this diagram makes a parallelogram. 
The roots $\vec{\a}_1$ and $\vec{\a}_3$ are orthogonal. Therefore elementary walls $\la 3\leftarrow 5  \ra$ and $\la 5\leftarrow 6  \ra$ are penetrable. Elementary walls $\la 3\leftarrow 4  \ra$ and $\la 4\leftarrow 6  \ra$ are also penetrable. The observable quantities $\mathrm{tr}\S$ and $T$ (\ref{eq:inv_quant}) of double wall $\la 3\leftarrow 5 \leftarrow 6 \ra$ are plotted in Figure \ref{eq:n4h356}.

Double wall $\la 2 \leftarrow  3  \leftarrow 5 \ra$ can be compressed. The moduli matrix is
\bea
H_{0\la 2 \leftarrow  3  \leftarrow 5 \ra}&=&H_{0\la 2 \ra}e^{E_2(r_1)}e^{E_1(r_2)}\nn\\
&=&\lt(
\begin{array}{cccc|cccc}
1 &  e^{r_2}  &    ~         &   ~   &   0           &   ~        &   ~  &   ~ \\
~ &  1        &    e^{r_1}   &   ~   &   ~           &   0        &   ~  &   ~ \\
~ &  ~        &    0         &   ~   &   e^{r_1+r_2} &   -e^{r_1} &   1  &   ~ \\
~ &  ~        &    ~         &   0   &   ~           &   ~        &   ~  &   1
\end{array}
\rt).
\eea
Under the worldvolume symmetry transformation, $H_{0\la 2 \leftarrow  3  \leftarrow 5 \ra}$
can be transformed to
\bea
H_{0\la 2 \leftarrow  3  \leftarrow 5 \ra}
&\rightarrow & 
\lt(
\begin{array}{cccc}
1   &   -e^{r_2}  &   ~   &   ~  \\
~   &   1         &   ~   &   ~  \\
~   &   ~         &   1   &   ~  \\
~   &   ~         &   ~   &   1  
\end{array}
\rt)\lt(
\begin{array}{cccc|cccc}
1 &  e^{r_2}  &    ~         &   ~   &   0           &   ~        &   ~  &   ~ \\
~ &  1        &    e^{r_1}   &   ~   &   ~           &   0        &   ~  &   ~ \\
~ &  ~        &    0         &   ~   &   e^{r_1+r_2} &   -e^{r_1} &   1  &   ~ \\
~ &  ~        &    ~         &   0   &   ~           &   ~        &   ~  &   1
\end{array}
\rt) \nn\\
&=&\lt(
\begin{array}{cccc|cccc}
1 &  ~        &    -e^{r_1+r_2}  &   ~   &   0           &   ~        &   ~  &   ~ \\
~ &  1        &    e^{r_1}     &   ~   &   ~           &   0        &   ~  &   ~ \\
~ &  ~        &    0           &   ~   &   e^{r_1+r_2} &   -e^{r_1} &   1  &   ~ \\
~ &  ~        &    ~           &   0   &   ~           &   ~        &   ~  &   1
\end{array}
\rt).\label{eq:h235}
\eea
In the limit where $r_1+r_2=r~(\mathrm{finite})$ and $r_1\rightarrow -\infty$, double wall $H_{0\la 2 \leftarrow  3  \leftarrow 5 \ra}$ becomes a compressed wall of level one $H_{0\la 2 \leftarrow  5 \ra}=H_{0\la 2 \ra}e^{[E_2,E_1](r)}$. 

\begin{figure}[ht!]
\begin{center}
$\begin{array}{cc}
\includegraphics[width=7.5cm,clip]{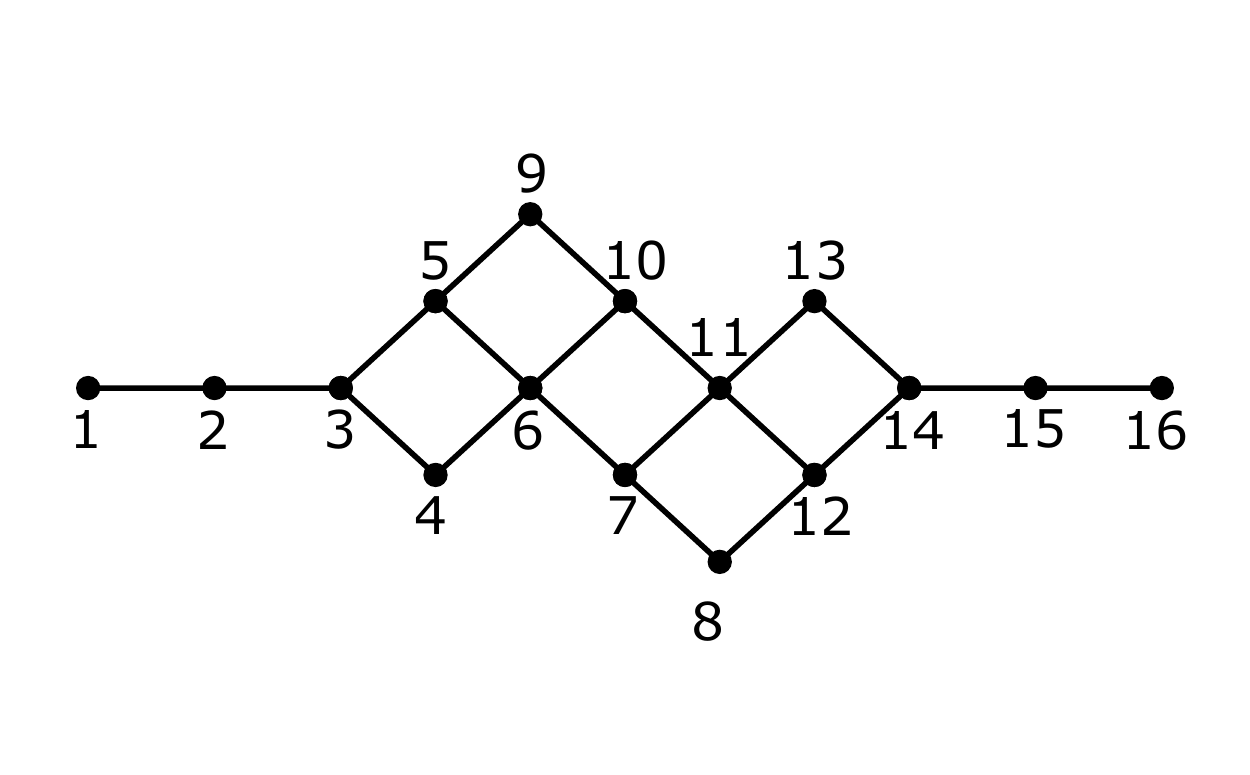}
&
\includegraphics[width=7.5cm,clip]{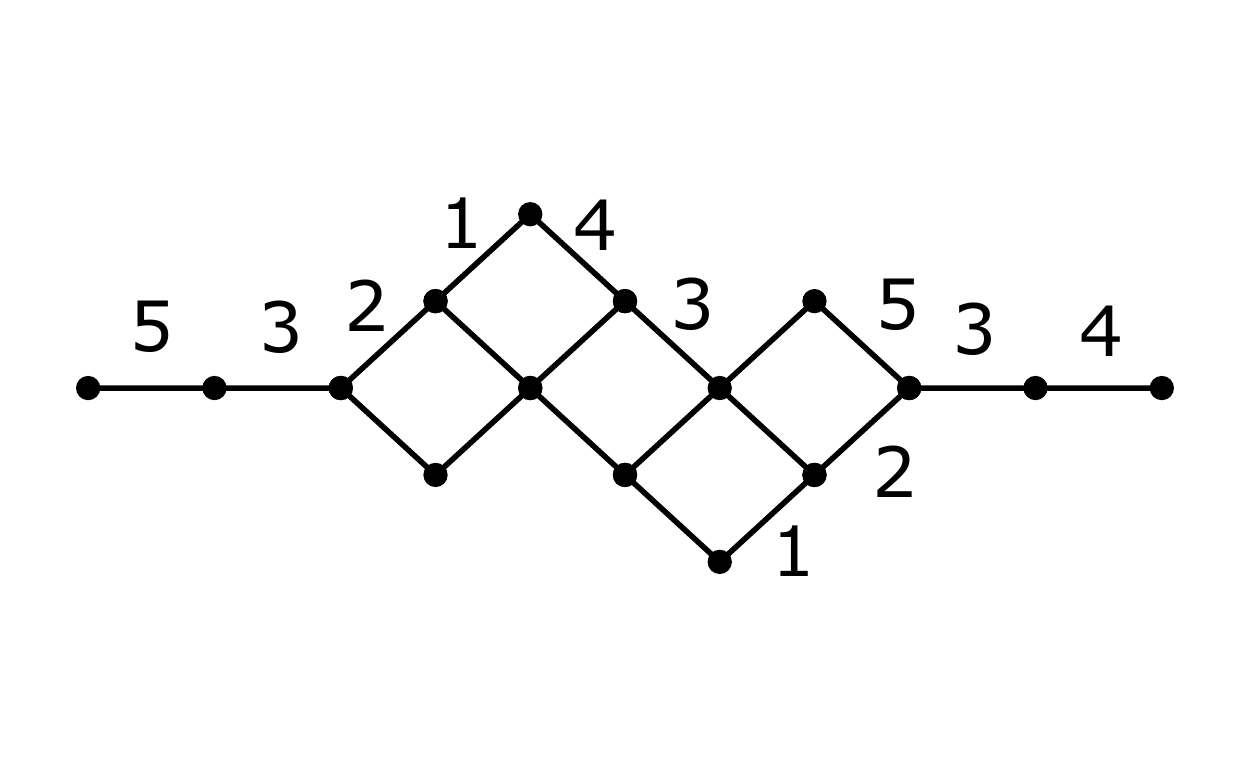}\\
\mathrm{(a)} & \mathrm{(b)}
\end{array}
$
\end{center}
 \caption{(a)Vacua of $SO(10)/U(5)$. The numbers indicate the labels of vacua. (b)Elementary walls of $SO(10)/U(5)$. The numbers indicate the subscript $i$ of simple roots $\vec{\a}_i$, $(i=1,\cdots,5)$. Since facing sides are the same, only one sides of each parallelogram are labelled. }
 \label{fig:n5}
\end{figure}

In $SO(10)/U(5)$, there are $16(=2^4)$ vacua. The five simple roots of $SO(10)$ are
\bea
&&\vec{\a}_1:=(1,-1,0,0,0), \nn\\
&&\vec{\a}_2:=(0,1,-1,0,0), \nn\\
&&\vec{\a}_3:=(0,0,1,-1,0), \nn\\
&&\vec{\a}_4:=(0,0,0,1,-1), \nn\\
&&\vec{\a}_5:=(0,0,0,1,1). 
\eea
The roots of elementary walls are
\bea
g_{\la 4 \leftarrow 5 \ra}=g_{\la 7 \leftarrow 8 \ra}
=g_{\la 9 \leftarrow 10 \ra}=g_{\la 12 \leftarrow 13 \ra}=\vec{\a}_1, \nn\\
g_{\la 3 \leftarrow 4 \ra}=g_{\la 6 \leftarrow 7 \ra}
=g_{\la 10 \leftarrow 11 \ra}=g_{\la 13 \leftarrow 14 \ra}=\vec{\a}_2, \nn\\
g_{\la 2 \leftarrow 3 \ra}=g_{\la 7 \leftarrow 9 \ra}
=g_{\la 8 \leftarrow 10 \ra}=g_{\la 14 \leftarrow 15 \ra}=\vec{\a}_3, \nn\\
g_{\la 3 \leftarrow 6 \ra}=g_{\la 4 \leftarrow 7 \ra}
=g_{\la 5 \leftarrow 8 \ra}=g_{\la 15 \leftarrow 16 \ra}=\vec{\a}_4, \nn\\
g_{\la 1 \leftarrow 2 \ra}=g_{\la 9 \leftarrow 12 \ra}
=g_{\la 10 \leftarrow 13 \ra}=g_{\la 11 \leftarrow 14 \ra}=\vec{\a}_5.
\eea
The roots of elementary walls are depicted in Figure \ref{fig:n5}. As before a pair of facing sides of each parallelogram are the same and a pair of adjacent sides of each parallelogram are orthogonal.
  
\begin{figure}[ht!]
\begin{center}
\includegraphics[height=9cm,clip]{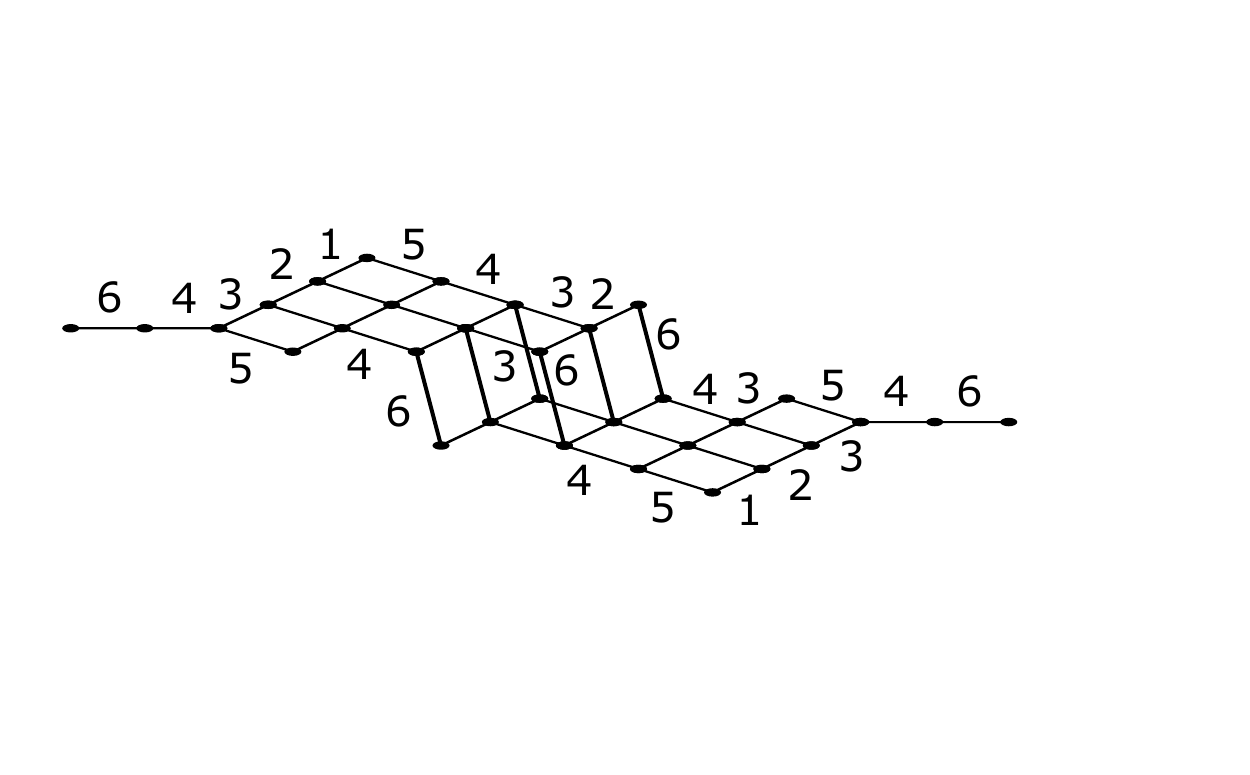}
\end{center}\vspace{-3cm}
 \caption{Elementary walls of $SO(12)/U(6)$. The numbers indicate the subscript $i$ of simple roots $\vec{\a}_i$, $(i=1,\cdots,6)$.}
 \label{fig:n6}
\end{figure}

\begin{figure}[ht!]
\begin{center}
\includegraphics[width=14cm,clip]{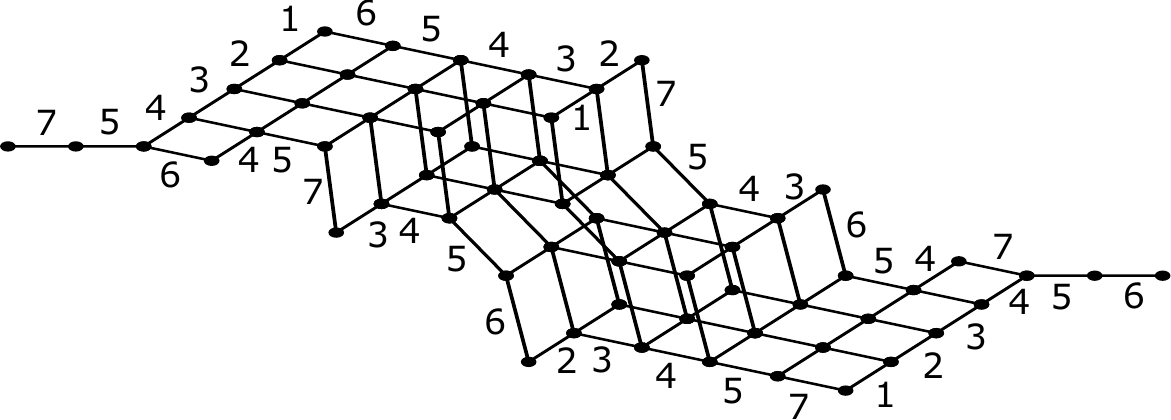}
\end{center}
 \caption{Elementary walls of $SO(14)/U(7)$. The numbers indicate the subscript $i$ of simple roots $\vec{\a}_i$, $(i=1,\cdots,7)$.}
 \label{fig:n7}
\end{figure}

The roots of elementary walls of $SO(12)/U(6)$ and $SO(14)/U(7)$ are depicted in Figure \ref{fig:n6} and Figure \ref{fig:n7}. We present the same diagrams labelled by the vacua in \ref{sec:app2} to avoid cluttering up pages.

\section{Generalization}\label{sec:Ngeneral}
The vacua are parametrized by $(\S_1,\cdots,\S_N)=(\pm m_1,\cdots,\pm m_N)$ as mentioned previously. The half of the vacua which differ by even numbers of minus signs belong to one nonlinear sigma model and the other half of the vacua belong  to the other nonlinear sigma model which is related by parity. Therefore there are $2^{N-1}$ vacua in the nonlinear sigma models on $SO(2N)/U(N)$. 

From the diagrams for $N=2,\cdots,7$, we make the following observations:
\begin{itemize}
\item $N=2$
\bea
\la 1 \ra &\leftarrow& \vec{2},\nn\\
\vec{2}~&\leftarrow& \la 2\ra.
\eea
\item $N=3$
\bea
\vec{3}\leftarrow \la 2 \ra \leftarrow \vec{1} \leftarrow \la 3 \ra 
\leftarrow \vec{2}. 
\eea
\item $N=4$
\bea
\vec{4}\leftarrow \vec{2}\leftarrow \la 3 \ra \leftarrow 
\lt\{\vec{1},\vec{3} \rt\}\leftarrow\cdots\leftarrow
\lt\{\vec{1},\vec{3} \rt\}
\leftarrow  \la 6 \ra  \leftarrow \vec{2} \leftarrow \vec{4}.
\eea
\item $N=5$
\bea
\vec{5}\leftarrow\cdots\leftarrow\lt\{
\vec{2},\vec{4}
\rt\} \leftarrow \la 6 \ra 
\leftarrow
\lt\{
\vec{1},\vec{3}
\rt\} \leftarrow \cdots \nn\\
\cdots\leftarrow
\lt\{
\vec{1},\vec{3}
\rt\} \leftarrow \la 11 \ra 
\leftarrow
\lt\{
\vec{2},\vec{5}
\rt\}\leftarrow\cdots\leftarrow \vec{4}.
\eea
\item $N=6$
\bea
\vec{6}\leftarrow\cdots\leftarrow\lt\{\vec{2},\vec{4}\rt\} \leftarrow \la 11 \ra 
\leftarrow
\lt\{\vec{1},\vec{3},\vec{6}
\rt\} \leftarrow\cdots \nn\\
\cdots\leftarrow \lt\{\vec{1},\vec{3},\vec{6}
\rt\} \leftarrow \la 22 \ra 
\leftarrow
\lt\{\vec{2},\vec{4}
\rt\}\leftarrow\cdots\leftarrow \vec{6}.
\eea
\item $N=7$
\bea
\vec{7}\leftarrow\cdots\leftarrow\lt\{\vec{2},\vec{4},\vec{7}\rt\} 
\leftarrow \la 22 \ra 
\leftarrow
\lt\{\vec{1},\vec{3},\vec{5}
\rt\} \leftarrow \cdots \nn\\
\cdots\leftarrow
\lt\{\vec{1},\vec{3},\vec{5}
\rt\} \leftarrow \la 43 \ra
\leftarrow
\lt\{\vec{2},\vec{4},\vec{6}
\rt\} \leftarrow \cdots \leftarrow \vec{6}.
\eea
\end{itemize}
The vacuum structures which are connected to the maximum number of simple roots are as follows.
\begin{itemize}
\item $N=4m-2$  $(m\geq 2)$
\bea
&&\vec{N}(=4m-2)\leftarrow\cdots \nn\\
&&\cdots\leftarrow\Big\{\underbrace{\vec{2},\vec{4},\cdots,\vec{4m-4}}_{(2m-2)}\Big\}\leftarrow \la A \ra \leftarrow \nn\\
&&\leftarrow
\Big\{\underbrace{\underbrace{\vec{1},\vec{3},\cdots,\vec{4m-5}}_{(2m-2)},\vec{4m-2}}_{(2m-1)}\Big\}\leftarrow\cdots \nn\\
&&\cdots\leftarrow\Big\{\underbrace{\vec{1},\vec{3},\cdots,\vec{4m-5},\vec{4m-2}}_{(2m-1)}\Big\}\leftarrow\la B \ra 
\leftarrow \nn\\
&&\leftarrow \Big\{\underbrace{\vec{2},\vec{4},\cdots,\vec{4m-4}}_{(2m-2)}\Big\}
\leftarrow\cdots\nn\\
&&\cdots\leftarrow \vec{N}(=4m-2). \label{eq:4m-2}
\eea
\item $N=4m-1$ $(m\geq 2)$
\bea
&&\vec{N}(=4m-1) \leftarrow \cdots \nn \\
&&\cdots\leftarrow\Big\{\underbrace{\underbrace{\vec{2},\vec{4},\cdots,\vec{4m-4}}_{(2m-2)},\vec{4m-1}}_{(2m-1)}\Big\}\leftarrow \la A \ra \leftarrow \nn\\
&&\leftarrow\Big\{\underbrace{\vec{1},\vec{3},\cdots,\vec{4m-5},\vec{4m-3}}_{(2m-1)}\Big\}\cdots \nn\\
&&\cdots\leftarrow\Big\{\underbrace{\vec{1},\vec{3},\cdots,\vec{4m-5},\vec{4m-3}}_{(2m-1)}\Big\}\leftarrow\la B \ra \leftarrow \nn\\
&&\leftarrow\Big\{\underbrace{\vec{2},\vec{4},\cdots,4m-2}_{(2m-1)}\Big\}\leftarrow\cdots \nn\\
&&\leftarrow \vec{N-1}(=4m-2). \label{eq:4m-1}
\eea
\item $N=4m$ $(m\geq 2)$
\bea
&&\vec{N}(=4m)\leftarrow\cdots \nn\\
&&\cdots\leftarrow\Big\{\underbrace{\vec{2},\vec{4},\cdots,\vec{4m-2}}_{(2m-1)}\Big\}\leftarrow\la A \ra \leftarrow \nn\\
&&\leftarrow\Big\{\underbrace{\vec{1},\vec{3},\cdots,\vec{4m-1}}_{(2m)}\Big\}\leftarrow\cdots  \nn\\
&&\cdots\leftarrow \Big\{\underbrace{\vec{1},\vec{3},\cdots,\vec{4m-1}}_{(2m)}\Big\}\leftarrow \la B \ra \leftarrow \nn\\
&&\leftarrow \Big\{\underbrace{\vec{2},\vec{4},\cdots,\vec{4m-2} }_{(2m-1)}\Big\} \leftarrow \cdots \nn\\
&&\cdots\leftarrow \vec{N}(=4m). \label{eq:4m}
\eea
\item $N=4m+1$ $(m\geq 2)$
\bea
&& \vec{N}(=4m+1)\leftarrow \cdots \nn\\
&&\cdots \leftarrow \Big\{\underbrace{\vec{2},\vec{4},\cdots,\vec{4m}}_{(2m)}\Big\}\leftarrow \la A \ra \leftarrow \nn\\
&&\leftarrow\Big\{\underbrace{\vec{1},\vec{3},\cdots,\vec{4m-1}}_{(2m)}\Big\}\leftarrow\cdots\nn\\
&&\cdots\leftarrow\Big\{\underbrace{\vec{1},\vec{3},\cdots,\vec{4m-1}}_{(2m)}\Big\}\leftarrow\la B \ra \leftarrow \nn\\
&&\leftarrow\Big\{\underbrace{\underbrace{\vec{2},\vec{4},\cdots,\vec{4m-2}}_{(2m-1)},\vec{4m+1}}_{(2m)}\Big\}\leftarrow
 \cdots\nn\\
&&\cdots\leftarrow \vec{N-1}(=4m). \label{eq:4m+1}
\eea
\end{itemize}
$\la A \ra$ and $\la B \ra$ are the vacua which are connected to the maximum number of elementary walls. $\la A \ra$ denotes the vacuum near $\la 1 \ra$ and $\la B \ra$ denotes the vacuum near $\la 2^{N-1} \ra$. (\ref{eq:4m-2}), (\ref{eq:4m-1}), (\ref{eq:4m}) and (\ref{eq:4m+1}) are proved in \ref{sec:app3}.

\section{Walls of nonlinear sigma model on $SO(12)/U(6)$}\label{sec:observation}
We have studied the vacuum structures which are connected to the maximum number of elementary walls for general $N$. Walls can be penetrable or compressed to a single wall. We discuss some physical consequences on $SO(12)/U(6)$, which is the simplest nontrival case. As it is shown previously, the vacua which are connected to the maximum number of elementary walls are $\la 11 \ra$ and $\la 22 \ra$. The structure near $\la 11 \ra$ is 
\bea
\begin{array}{ccccc}
\begin{array}{ccc}
\la 7 \ra & \leftarrow & \vec{2} \\
\la 10 \ra & \leftarrow & \vec{4}
\end{array}
& \leftarrow & \la 11 \ra & \leftarrow &
\begin{array}{ccc}
 \vec{1} & \leftarrow & \la 19 \ra \\
 \vec{3} & \leftarrow & \la 13 \ra \\
 \vec{6} & \leftarrow & \la 12 \ra
\end{array}
\end{array} \label{eq:n6_vac11}
\eea

\begin{figure}[ht!]
\vspace{2cm}
\begin{center}
$\begin{array}{ccc}
\includegraphics[width=5cm,clip]{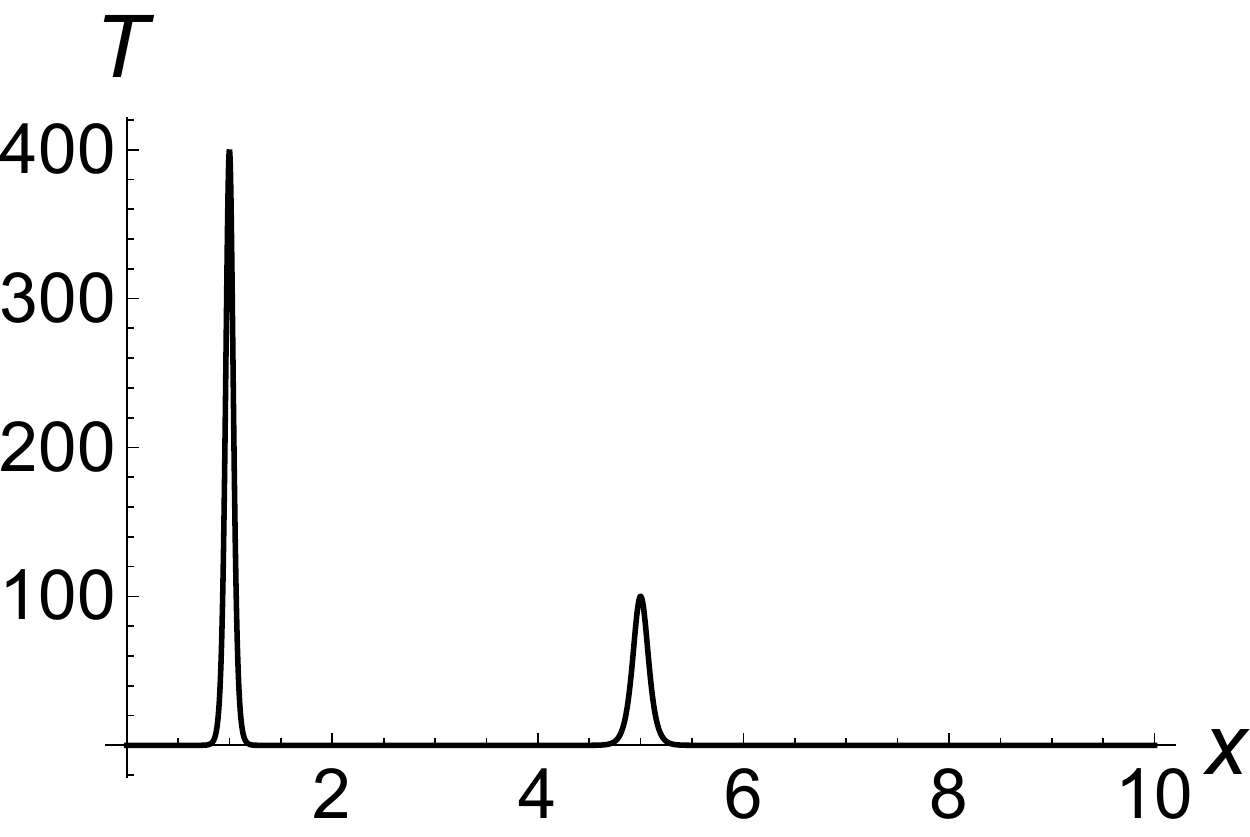}
&
\includegraphics[width=5cm,clip]{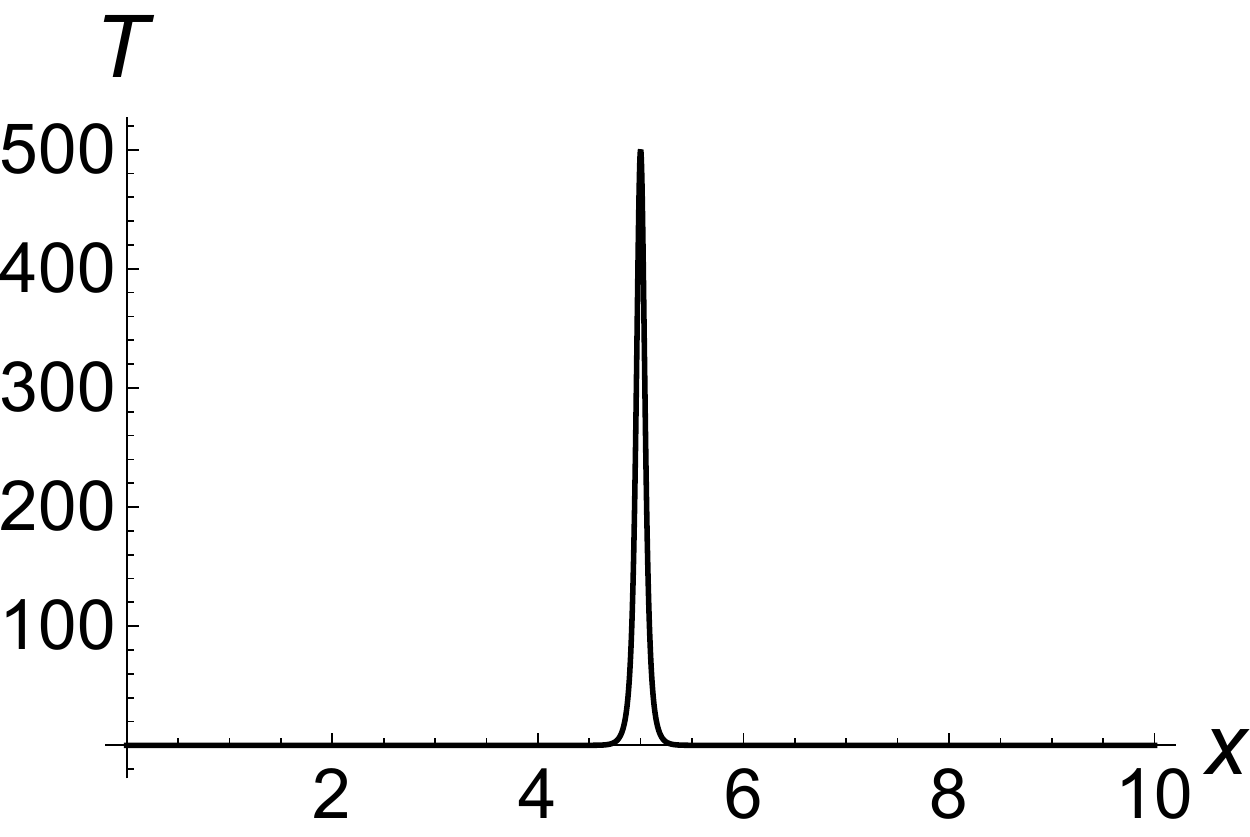}
& 
\includegraphics[width=5cm,clip]{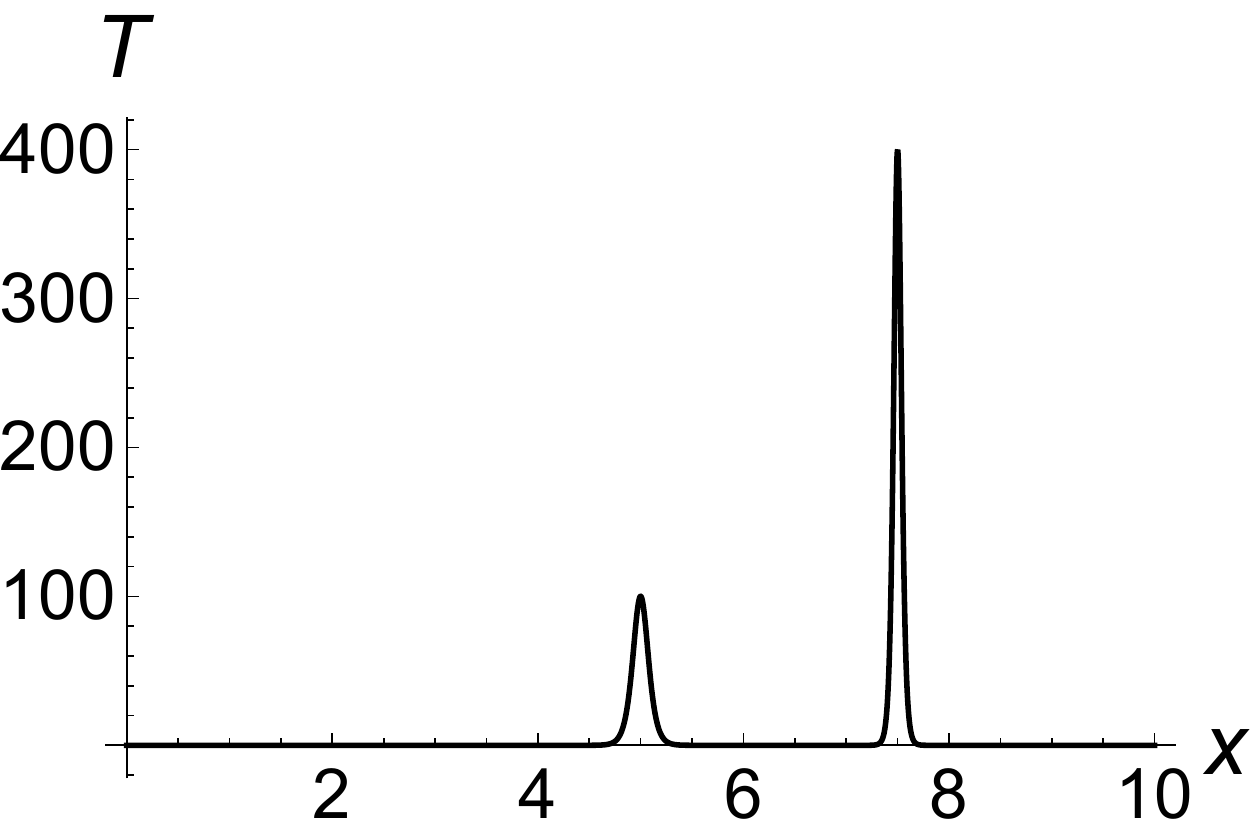} \\
\mathrm{(a)} & \mathrm{(b)} & \mathrm{(c)}
\end{array}
$
\end{center}
 \caption{Double wall $\la 10 \leftarrow  11  \leftarrow 19 \ra$ in $SO(12)/U(6)$, which consists of two penetrable walls. 
 $m_1=80$, $m_2=60$, $m_3=40$, $m_4=20$, $m_5=10$, $m_6=5$. (a)$r_1=50$, $r_2=20$ (b)$r_1=50$, $r_2=100$ (c)$r_1=50$, $r_2=150$.}
 \label{fig:n6h101119}
\end{figure}

In (\ref{eq:n6_vac11}), $\vec{\a_4}\cdot\vec{\a_1}=0$. Therefore the elementary wall which interpolates $\la 10 \ra$ and $\la 11 \ra$ and the elementary wall which interpolates $\la 11 \ra$ and $\la 19 \ra$ are penetrable. The moduli matrix of the double wall which connects $\la 10 \ra$, $\la 11 \ra$ and $\la 19 \ra$ is
\bea
H_{0\la 10 \leftarrow 11 \leftarrow 19 \ra}
&=&H_{0\la 10 \ra}e^{E_4(r_1)}e^{E_1(r_2)}  \nn\\
&=&\lt(
\begin{array}{cccccc|cccccc}
1 & e^{r_2} & 0 & 0 & 0 & 0 & 0 & 0 & 0 & 0 & 0 & 0  \\
0 & 0 & 0 & 0 & 0 & 0 & -e^{r_2} & 1 & 0 & 0 & 0 & 0  \\
0 & 0 & 1 & 0 & 0 & 0 & 0 & 0 & 0 & 0 & 0 & 0  \\
0 & 0 & 0 & 1 & e^{r_1} & 0 & 0 & 0 & 0 & 0 & 0 & 0  \\
0 & 0 & 0 & 0 & 0 & 0 & 0 & 0 & 0 & -e^{r_1} & 1 & 0  \\
0 & 0 & 0 & 0 & 0 & 1 & 0 & 0 & 0 & 0 & 0 & 0  \\
\end{array}
\rt).
\eea
The tension of the walls are plotted in Figure \ref{fig:n6h101119}. Elementary wall $\la 10 \leftarrow 11 \ra$ and elementary wall $\la 11 \leftarrow 19 \ra$ pass through each other.

\begin{figure}[ht!]
\vspace{2cm}
\begin{center}
$\begin{array}{ccc}
\includegraphics[width=5cm,clip]{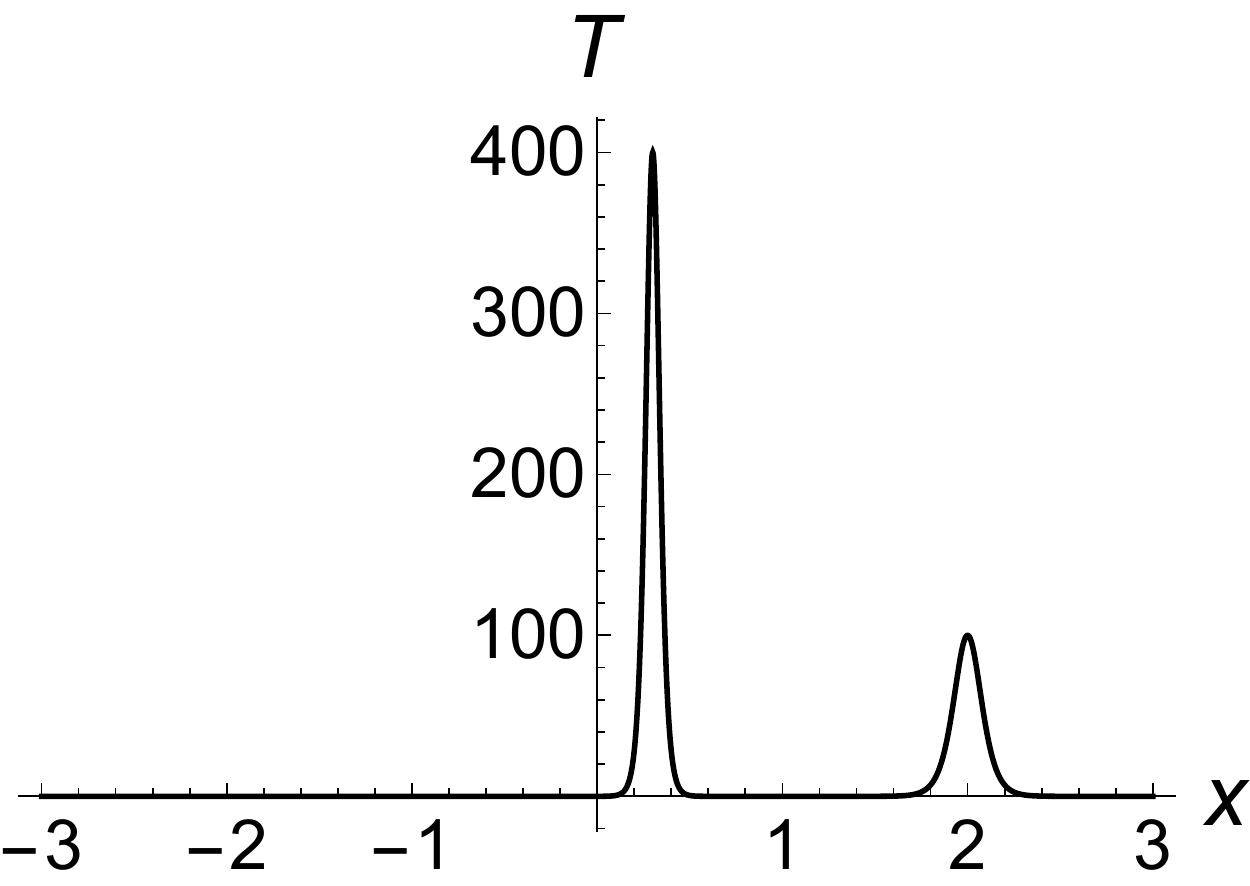}
&
\includegraphics[width=5cm,clip]{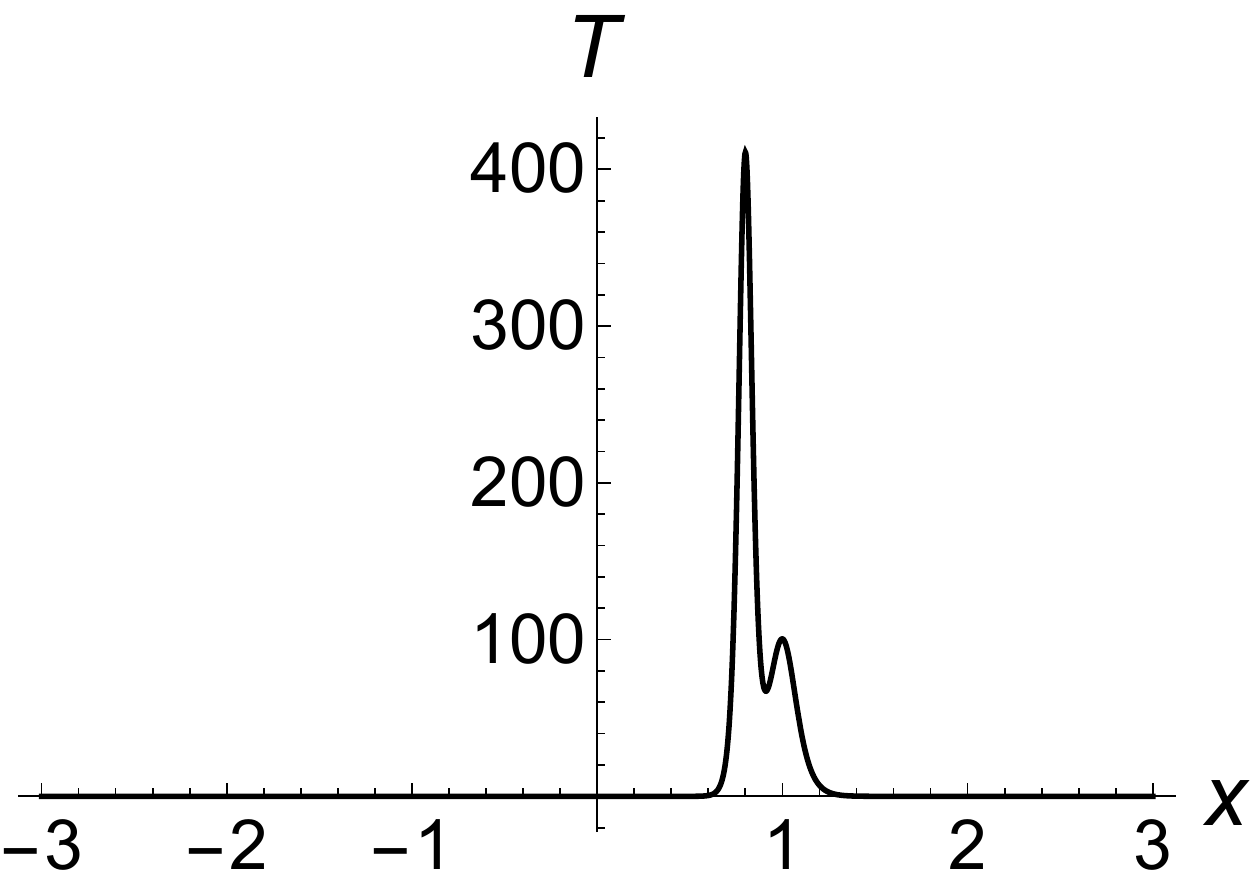}
& 
\includegraphics[width=5cm,clip]{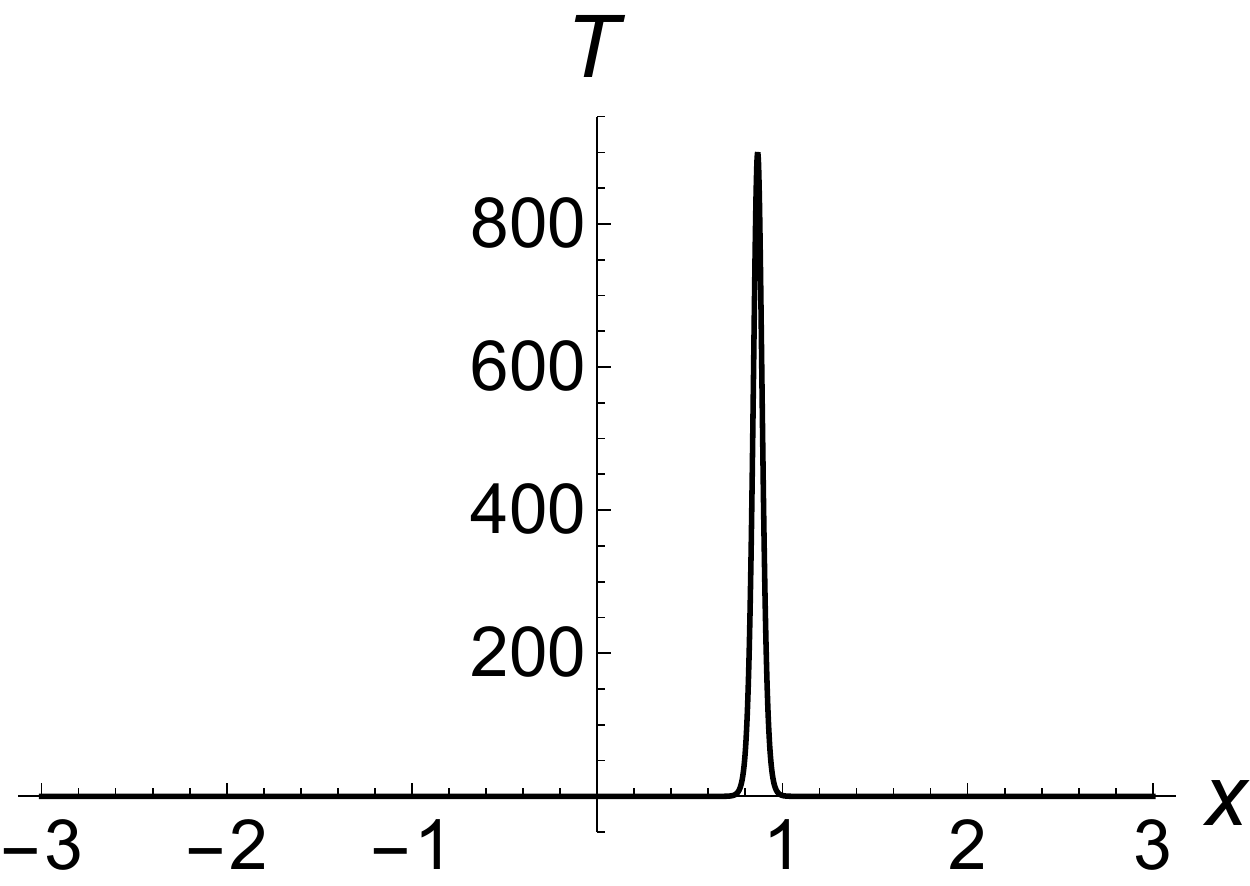} \\
\mathrm{(a)} & \mathrm{(b)} & \mathrm{(c)}
\end{array}
$
\end{center}
 \caption{Double wall $\la 10 \leftarrow  11  \leftarrow 13 \ra$ in $SO(12)/U(6)$, which consists of two penetrable walls. 
 $m_1=80$, $m_2=60$, $m_3=40$, $m_4=20$, $m_5=10$, $m_6=5$. (a)$r_1=20$, $r_2=6$ (b)$r_1=10$, $r_2=16$ (c)$r_1=-10$, $r_2=36$.}
 \label{fig:n6h101113}
\end{figure}

In (\ref{eq:n6_vac11}), $\vec{\a_4}\cdot\vec{\a_3}\neq0$. Therefore elementary wall $\la 10 \leftarrow 11 \ra$ and elementary wall $\la 11 \leftarrow 13 \ra$ can be compressed to a single wall. The moduli matrix of double wall $\la 10 \leftarrow 11 \leftarrow 13 \ra$ is
\bea
H_{0\la 10 \leftarrow 11 \leftarrow 13 \ra}
&=&H_{0\la 10 \ra}e^{E_4(r_1)}e^{E_3(r_2)}  \nn\\
&=&\lt(
\begin{array}{cccccc|cccccc}
1 & 0 & 0 & 0 & 0 & 0 & 0 & 0 & 0 & 0 & 0 & 0  \\
0 & 0 & 0 & 0 & 0 & 0 & 0 & 1 & 0 & 0 & 0 & 0  \\
0 & 0 & 1 & e^{r_2} & 0 & 0 & 0 & 0 & 0 & 0 & 0 & 0  \\
0 & 0 & 0 & 1 & e^{r_1} & 0 & 0 & 0 & 0 & 0 & 0 & 0  \\
0 & 0 & 0 & 0 & 0 & 0 & 0 & 0 & e^{r_1+r_2} & -e^{r_1} & 1 & 0  \\
0 & 0 & 0 & 0 & 0 & 1 & 0 & 0 & 0 & 0 & 0 & 0  \\
\end{array}
\rt).
\eea
By using the worldvolume symmetry the moduli matrix can be transformed to 
\bea
&&H_{0\la 10 \leftarrow 11 \leftarrow 13 \ra} \nn\\
&&\rightarrow
\lt(
\begin{array}{cccccc}
1 & 0 & 0 & 0 & 0 & 0   \\
0 & 1 & 0 & 0 & 0 & 0   \\
0 & 0 & 1 & -e^{r_2} & 0 & 0   \\
0 & 0 & 0 & 1 & 0 & 0  \\
0 & 0 & 0 & 0 & 1 & 0  \\
0 & 0 & 0 & 0 & 0 & 1 \\
\end{array}
\rt)
\lt(
\begin{array}{cccccc|cccccc}
1 & 0 & 0 & 0 & 0 & 0 & 0 & 0 & 0 & 0 & 0 & 0  \\
0 & 0 & 0 & 0 & 0 & 0 & 0 & 1 & 0 & 0 & 0 & 0  \\
0 & 0 & 1 & e^{r_2} & 0 & 0 & 0 & 0 & 0 & 0 & 0 & 0  \\
0 & 0 & 0 & 1 & e^{r_1} & 0 & 0 & 0 & 0 & 0 & 0 & 0  \\
0 & 0 & 0 & 0 & 0 & 0 & 0 & 0 & e^{r_1+r_2} & -e^{r_1} & 1 & 0  \\
0 & 0 & 0 & 0 & 0 & 1 & 0 & 0 & 0 & 0 & 0 & 0  \\
\end{array}
\rt) \nn \\
&&=
\lt(
\begin{array}{cccccc|cccccc}
1 & 0 & 0 & 0 & 0 & 0 & 0 & 0 & 0 & 0 & 0 & 0  \\
0 & 0 & 0 & 0 & 0 & 0 & 0 & 1 & 0 & 0 & 0 & 0  \\
0 & 0 & 1 & 0 & -e^{r_1+r_2} & 0 & 0 & 0 & 0 & 0 & 0 & 0  \\
0 & 0 & 0 & 1 & e^{r_1} & 0 & 0 & 0 & 0 & 0 & 0 & 0  \\
0 & 0 & 0 & 0 & 0 & 0 & 0 & 0 & e^{r_1+r_2} & -e^{r_1} & 1 & 0  \\
0 & 0 & 0 & 0 & 0 & 1 & 0 & 0 & 0 & 0 & 0 & 0  \\
\end{array}
\rt).
\eea
In the limit where $r_1+r_2=r$ (finite) and $r_1\rightarrow -\infty$, double wall $\la 10 \leftarrow 11 \leftarrow 13 \ra$ becomes a compressed wall of level one. The tension of double wall $\la 10 \leftarrow 11 \leftarrow 13 \ra$ is plottd in Figure \ref{fig:n6h101113}. Two elementary walls are compressed to a single wall.

\section{Discussion}\label{sec:discuss}
We have discussed the vacua and the walls of mass-deformed K\"{a}hler nonlinear sigma models on $SO(2N)/U(N)$ with $N\geq 2$ by using the moduli matrix formalism and the simple roots of $SO(2N)$. We have observed that there are penetrable walls in the cases of $N>3$, which are non-Abelian theories. We have discussed the vacuum structures which are connected to the maximum number of elementary walls and proved them by induction.

\vspace{1cm}

{\bf Acknowledgement}

We would like to thank M. Arai, V. Dobrev, M. Nitta, K. Ohashi, M. Park, N. Sakai and A. Zheltukhin for kind correspondences and helpful comments. BHL is supported by Basic Science Research Program through the National Research Foundation of Korea (NRF-2017R1D1A1B03028310). CP and SS are supported by the Korea Ministry of Education, Science and Technology, GyeongsangbukDo
and Pohang City. CP is also supported by Basic Science Research Program through the National
Research Foundation of Korea funded by the Ministry of Education (NRF-2016R1D1A1B03932371). SS is also supported by Basic Science Research Program through the National Research Foundation of Korea (NRF-2017R1D1A1B03034222).

%
%
\appendix
\def\thesection{Appendix \Alph{section}}
\setcounter{equation}{0}
\renewcommand{\theequation}{\Alph{section}.\arabic{equation}}
\section{Vacuum structure of nonlinear sigma models on $SO(2N)/U(N)$ with $N=2,\,3$}\label{sec:app1}
The vacua of nonlinear sigma models are obtained from the vacuum condition (\ref{eq:vac}) in Section \ref{sec:sec2}. 

The vacua of nonlinear sigma model on $SO(4)/U(2)$ are
\bea
&&\Phi_1=\lt(
\begin{array}{cc|cc}
1  &  ~ &  0  &  ~  \\
~  &  1 &  ~  &  0  
\end{array}
\rt),~~~(\S_1,\S_2)=(m_1,m_2),  \nn\\
&&\Phi_2=\lt(
\begin{array}{cc|cc}
0  &  ~ &  1  &  ~  \\
~  &  0 &  ~  &  1  
\end{array}
\rt),~~~(\S_1,\S_2)=(-m_1,-m_2), \nn\\
&&\Phi_3=\lt(
\begin{array}{cc|cc}
1  &  ~ &  0  &  ~  \\
~  &  0 &  ~  &  1  
\end{array}
\rt),~~~(\S_1,\S_2)=(m_1,-m_2),  \nn\\
&&\Phi_4=\lt(
\begin{array}{cc|cc}
0  &  ~ &  1  &  ~  \\
~  &  1 &  ~  &  0  
\end{array}
\rt),~~~(\S_1,\S_2)=(-m_1,m_2). 
\eea 
There are four solutions to the vacuum condition (\ref{eq:vac}), but only the half of them are the vacua of a nonlinear sigma model on $SO(4)/U(2)$ and the other half are the vacua of the other nonlinear sigma model which is related by a parity transformation. Let us define a rotation transformation $R$ and a parity transformation $P$ as follows:
\bea
R=\lt(
\begin{array}{cc|cc}
0  &  ~  &  1  &  ~  \\
~  &  0  &  ~  &  1  \\ \hline
1  &  ~  &  0  &  ~  \\
~  &  1  &  ~  &  0
\end{array}
\rt),~~~
P=\lt(
\begin{array}{cc|cc}
1  &  ~  &  0  &  ~  \\
~  &  0  &  ~  &  1  \\ \hline
0  &  ~  &  1  &  ~  \\
~  &  1  &  ~  &  0
\end{array}
\rt).
\eea
The vacua are related by
\bea
&&\Phi_2=\Phi_1\cdot R,~~\Phi_4=\Phi_3\cdot R,  \nn\\
&&\Phi_3=\Phi_1\cdot P,~~\Phi_4=\Phi_2\cdot P.
\eea
The vacua $\Phi_1$ and $\Phi_2$ are on the same manifold. $\Phi_3$ and $\Phi_4$ are on the other manifold which are related by the parity transformation $P$. Therefore we can focus only on $\Phi_1$ and $\Phi_2$ without loss of generality. The corresponding moduli matrices of the vacua, which are related to the vacuum solution by (\ref{eq:bps_sol}) are
\bea
&&H_{0\la 1 \ra}=\lt(
\begin{array}{cc|cc}
1   &  ~   &   0   &   ~   \\
~   &  1   &   ~   &   0 
\end{array}
\rt),~~(\S_1,\S_2)=(m_1,m_2), \nn\\
&&H_{0\la 2 \ra}=\lt(
\begin{array}{cc|cc}
0   &  ~   &   1   &   ~   \\
~   &  0   &   ~   &   1 
\end{array}
\rt),~~(\S_1,\S_2)=(-m_1,-m_2). \label{eq:modulimvac_2}
\eea
The vacua of nonlinear sigma model on $SO(6)/U(3)$ are
\begin{eqnarray}
&&\Phi_1=\left(
\begin{array}{ccc|ccc}
1   &  ~  &  ~  &  0  & ~ & ~ \\
~   &  1  &  ~  &  ~  & 0 & ~ \\
~   &  ~  &  1  &  ~  & ~ & 0
\end{array}
\right),~~(\S_1,\S_2,\S_3)=(m_1,m_2,m_3), \nn\\
&&\Phi_2=\left(
\begin{array}{ccc|ccc}
1   &  ~  &  ~  &  0  & ~ & ~ \\
~   &  0  &  ~  &  ~  & 1 & ~ \\
~   &  ~  &  0  &  ~  & ~ & 1
\end{array}
\right),~~(\S_1,\S_2,\S_3)=(m_1,-m_2,-m_3),\nn\\
&&\Phi_3=\left(
\begin{array}{ccc|ccc}
0   &  ~  &  ~  &  1  & ~ & ~ \\
~   &  1  &  ~  &  ~  & 0 & ~ \\
~   &  ~  &  0  &  ~  & ~ & 1
\end{array}
\right),~~(\S_1,\S_2,\S_3)=(-m_1,m_2,-m_3),\nn\\
&&\Phi_4=\left(
\begin{array}{ccc|ccc}
0   &  ~  &  ~  &  1  & ~ & ~ \\
~   &  0  &  ~  &  ~  & 1 & ~ \\
~   &  ~  &  1  &  ~  & ~ & 0
\end{array}
\right),~~(\S_1,\S_2,\S_3)=(-m_1,-m_2,m_3),\nn\\
&&\Phi_5=\left(
\begin{array}{ccc|ccc}
1   &  ~  &  ~  &  0  & ~ & ~ \\
~   &  1  &  ~  &  ~  & 0 & ~ \\
~   &  ~  &  0  &  ~  & ~ & 1
\end{array}
\right),~~(\S_1,\S_2,\S_3)=(m_1,m_2,-m_3),\nn\\
&&\Phi_6=\left(
\begin{array}{ccc|ccc}
1   &  ~  &  ~  &  0  & ~  & ~ \\
~   &  0  &  ~  &  ~  & 1  & ~ \\
~   &  ~  &  1  &  ~  & ~  & 0
\end{array}
\right),~~(\S_1,\S_2,\S_3)=(m_1,-m_2,m_3),\nn\\
&&\Phi_7=\left(
\begin{array}{ccc|ccc}
0   &  ~  &  ~  &  1  & ~ & ~ \\
~   &  1  &  ~  &  ~  & 0 & ~ \\
~   &  ~  &  1  &  ~  & ~ & 0
\end{array}
\right),~~(\S_1,\S_2,\S_3)=(-m_1,m_2,m_3),\nn\\
&&\Phi_8=\left(
\begin{array}{ccc|ccc}
0   &  ~  &  ~  &  1  & ~ & ~ \\
~   &  0  &  ~  &  ~  & 1 & ~ \\
~   &  ~  &  0  &  ~  & ~ & 1
\end{array}
\right),~~(\S_1,\S_2,\S_3)=(-m_1,-m_2,-m_3).
\end{eqnarray}
There are eight vacua, but only the half of them are the vacua of a nonlinear sigma model on $SO(6)/U(3)$. We can identify them by rotational transformations. We have three rotational transformations
\begin{eqnarray}
&&R_1=\lt(
\begin{array}{ccc|ccc}
1  &  ~  &  ~  &  0   &  ~ &  ~  \\
~  &  0  &  ~  &  ~   &  1 &  ~ \\
~  &  ~  &  0  &  ~   &  ~ &  1 \\ \hline
0  &  ~  &  ~  &  1   &  ~ &  ~  \\
~  &  1  &  ~  &  ~   &  0 &  ~ \\
~  &  ~  &  1  &  ~   &  ~ &  0 \\
\end{array}
\rt),~ R_2=\lt(
\begin{array}{ccc|ccc}
0  &  ~  &  ~  &  1   &  ~ &  ~  \\
~  &  1  &  ~  &  ~   &  0 &  ~ \\
~  &  ~  &  0  &  ~   &  ~ &  \\  \hline
1  &  ~  &  ~  &  0   &  ~ &  ~  \\
~  &  0  &  ~  &  ~   &  1 &  ~ \\
~  &  ~  &  1  &  ~   &  ~ &  1 \\
\end{array}
\rt), \nn\\
&& R_3=\lt(
\begin{array}{ccc|ccc}
0  &  ~  &  ~  &  1   &  ~ &  ~  \\
~  &  0  &  ~  &  ~   &  1 &  ~ \\
~  &  ~  &  1  &  ~   &  ~ &  0 \\ \hline
1  &  ~  &  ~  &  0   &  ~ &  ~  \\
~  &  1  &  ~  &  ~   &  0 &  ~ \\
~  &  ~  &  0  &  ~   &  ~ &  1 \\
\end{array}
\rt).
\end{eqnarray}
The vacua are related by
\begin{eqnarray}
\Phi_2=\Phi_1\cdot R_1,~~\Phi_3=\Phi_1\cdot R_2,~~\Phi_4=\Phi_1\cdot
R_3.
\end{eqnarray}
The solutions $\Ph_1$, $\Ph_2$, $\Ph_3$ and $\Ph_4$ are the vacua of a nonlinear sigma model on $SO(6)/U(3)$ and the others are the vacua of a nonlinear sigma model on the other $SO(6)/U(3)$ which are related by parity. The moduli matrices of the vacua are
\begin{eqnarray}
&&H_{0\la 1\ra}=\left(
\begin{array}{ccc|ccc}
1   &  ~  &  ~  &  0  & ~ & ~ \\
~   &  1  &  ~  &  ~  & 0 & ~ \\
~   &  ~  &  1  &  ~  & ~ & 0
\end{array}
\right),~~~(\S_1,\S_2,\S_3)=(m_1,m_2,m_3), \nn \\
&&H_{0\la 2\ra}=\left(
\begin{array}{ccc|ccc}
1   &  ~  &  ~  &  0  & ~ & ~ \\
~   &  0  &  ~  &  ~  & 1 & ~ \\
~   &  ~  &  0  &  ~  & ~ & 1
\end{array}
\right),~~~(\S_1,\S_2,\S_3)=(m_1,-m_2,-m_3)  \nn \\
&&H_{0\la 3\ra}=\left(
\begin{array}{ccc|ccc}
0   &  ~  &  ~  &  1  & ~ & ~ \\
~   &  1  &  ~  &  ~  & 0 & ~ \\
~   &  ~  &  0  &  ~  & ~ & 1
\end{array}
\right),~~~(\S_1,\S_2,\S_3)=(-m_1,m_2,-m_3) \nn \\
&&H_{0\la 4\ra}=\left(
\begin{array}{ccc|ccc}
0   &  ~  &  ~  &  1  & ~ & ~ \\
 ~   &  0  &  ~  &  ~  & 1 & ~ \\
~   &  ~  &  1  &  ~  & ~ & 0
\end{array}
\right),~~~(\S_1,\S_2,\S_3)=(-m_1,-m_2,m_3). \label{eq:modulimvac_3}
\end{eqnarray}
We use the same method to compute moduli matrices of vacua for any $N$. All the vacua are labelled by sets of $(\S_1,\cdots,\S_N)=(\pm m_1,\cdots,\pm m_N)$ of which the numbers of minus signs differ by even numbers.


\section{Vacua}\label{sec:app2}
\begin{figure}[ht!]
\begin{center}
\includegraphics[height=9cm,clip]{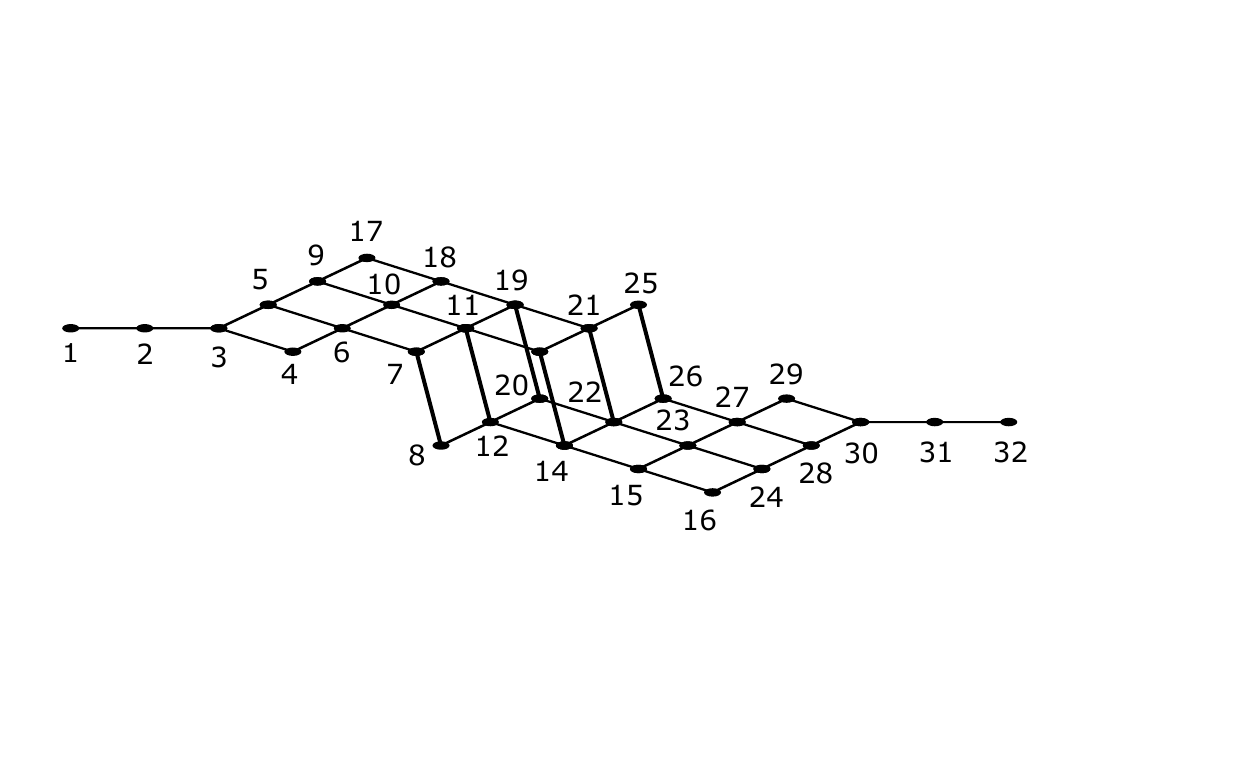}
\end{center}\vspace{-3cm}
 \caption{Vacua of $SO(12)/U(6)$. The numbers indicate the labels of vacua.}
 \label{fig:n6v}
\end{figure}

\begin{figure}[ht!]
\begin{center}
\includegraphics[width=14cm,clip]{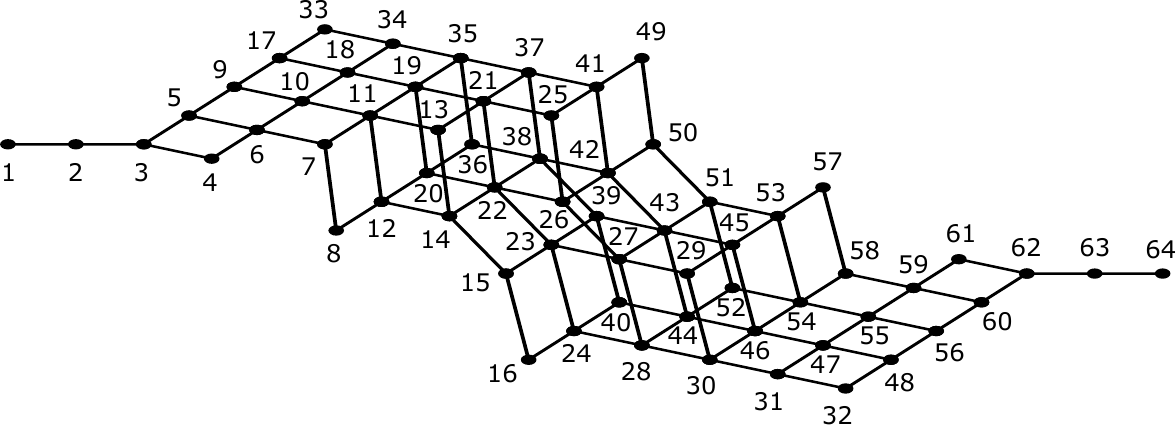}
\end{center}
 \caption{Vacua of $SO(14)/U(7)$. The numbers indicate the labels of vacua.}
 \label{fig:n7v}
\end{figure}

\section{Vaccuum structures}\label{sec:app3}
We prove (\ref{eq:4m-2}), (\ref{eq:4m-1}), (\ref{eq:4m}) and (\ref{eq:4m+1}). The vacua which are connected to the maximum number of elementary walls can be found by decomposing the diagrams into two-dimensional diagrams. The only rule is that simple roots which already have appeared in the previous diagrams should not be repeated.

\begin{figure}[ht!]
\begin{center}
$\begin{array}{ccc}
\includegraphics[width=6cm,clip]{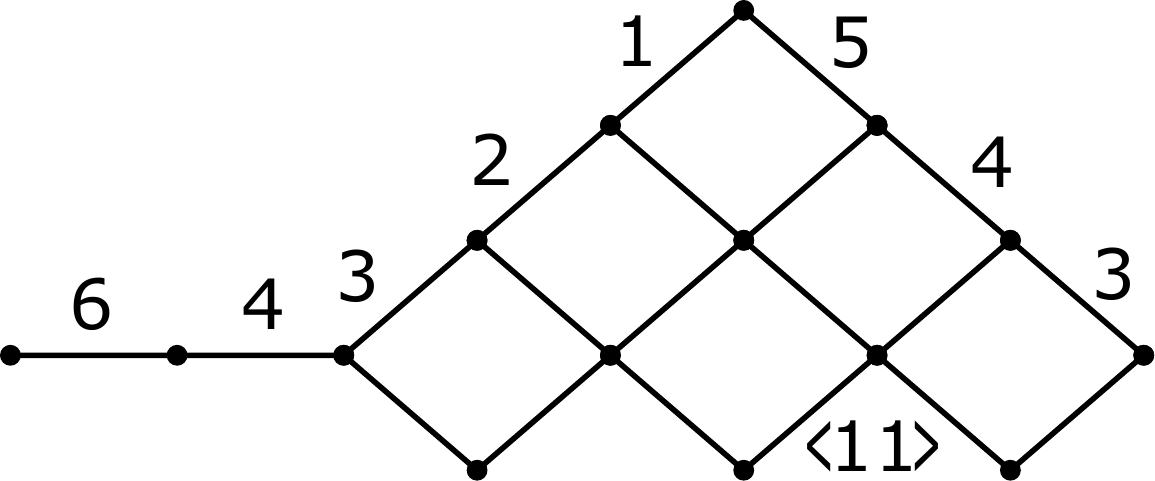}
&~~~~~~&
\includegraphics[width=3cm,clip]{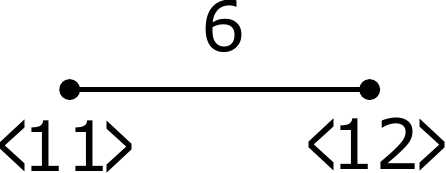}\vspace{1cm}\\
\mathrm{(a)} &~~~~~~& \mathrm{(b)}
\end{array}
$
\end{center}
 \caption{$N=6$ case. The vacuum structure near $\la 1 \ra$ decomposes into two diagrams. Diagram (b) is the same as Figure \ref{fig:n2n3rts} (a).}
 \label{fig:n6_decomp}
\end{figure}

\begin{figure}[ht!]
\begin{center}
$\begin{array}{ccc}
\includegraphics[width=7cm,clip]{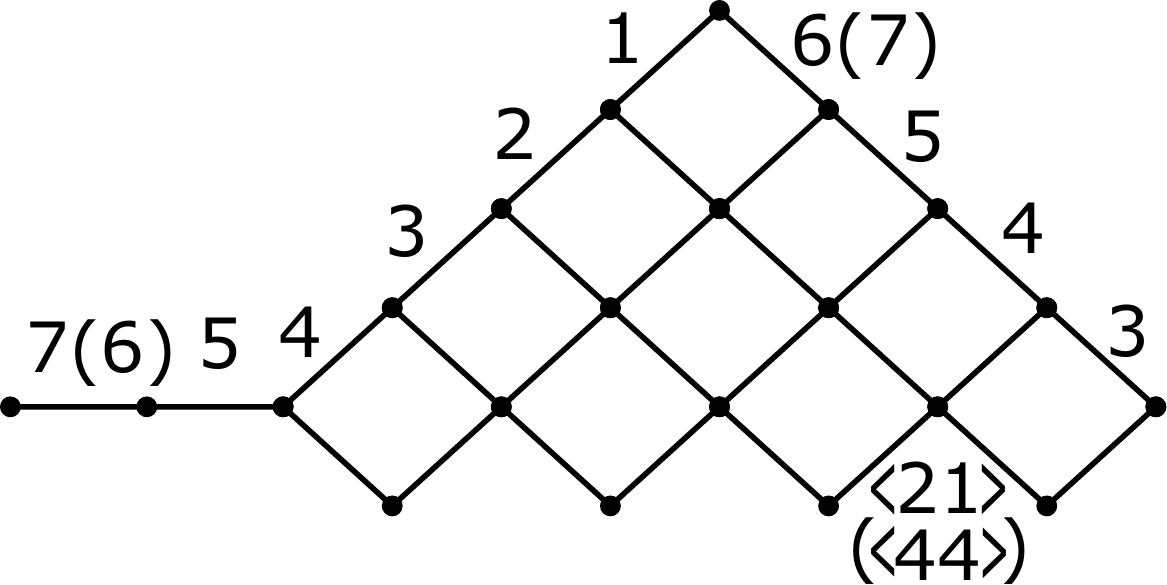}
&~~~~~~&
\includegraphics[width=6cm,clip]{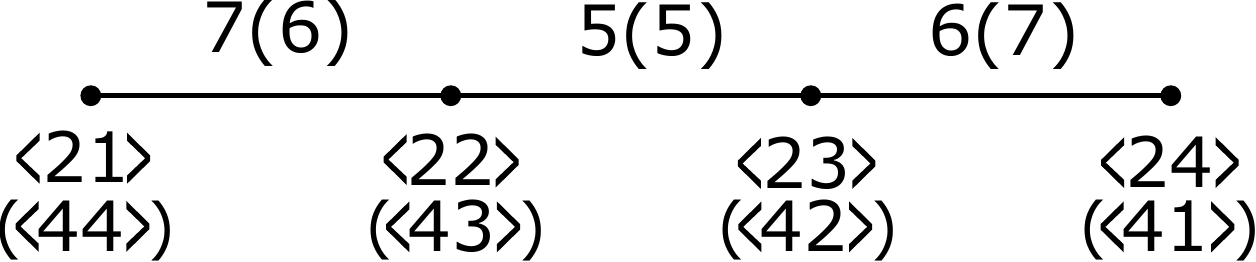}\vspace{1cm}\\
\mathrm{(a)} &~~~~~~& \mathrm{(b)}
\end{array}
$
\end{center}
\caption{$N=7$ case. Diagrams near $\la 1 \ra$($\la 64\ra$). Diagram (b) is the same as Figure \ref{fig:n2n3rts} (b).}
 \label{fig:n7_decomp}
\end{figure}

\begin{figure}[ht!]
\begin{center}
$\begin{array}{ccc}
\includegraphics[width=7cm,clip]{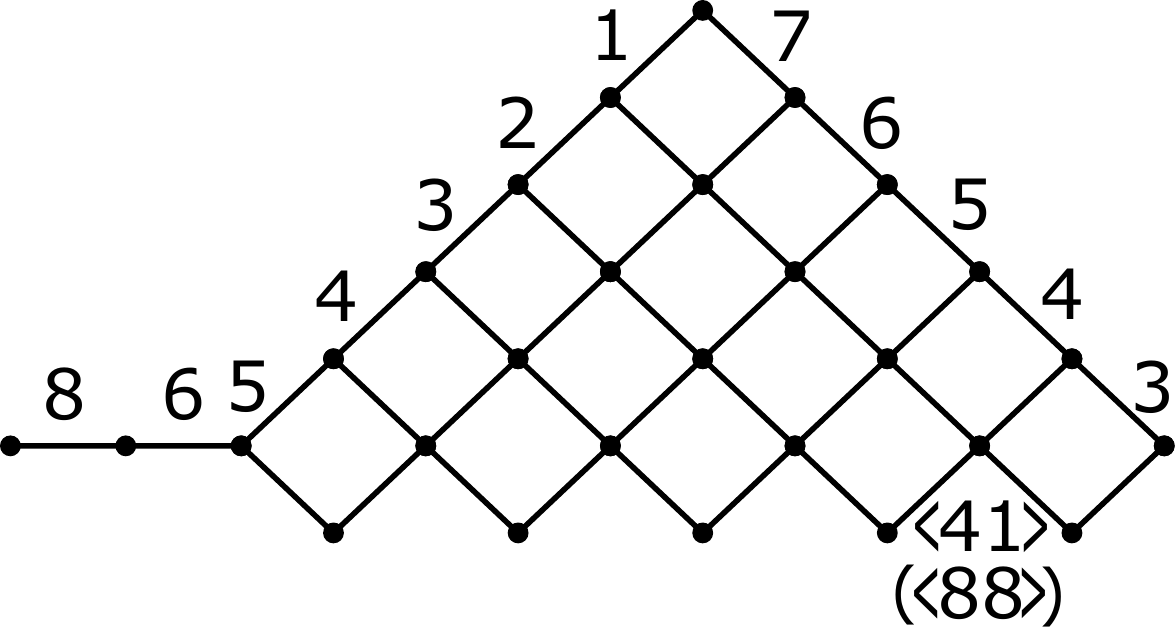}
&~~~~~~&
\includegraphics[width=7cm,clip]{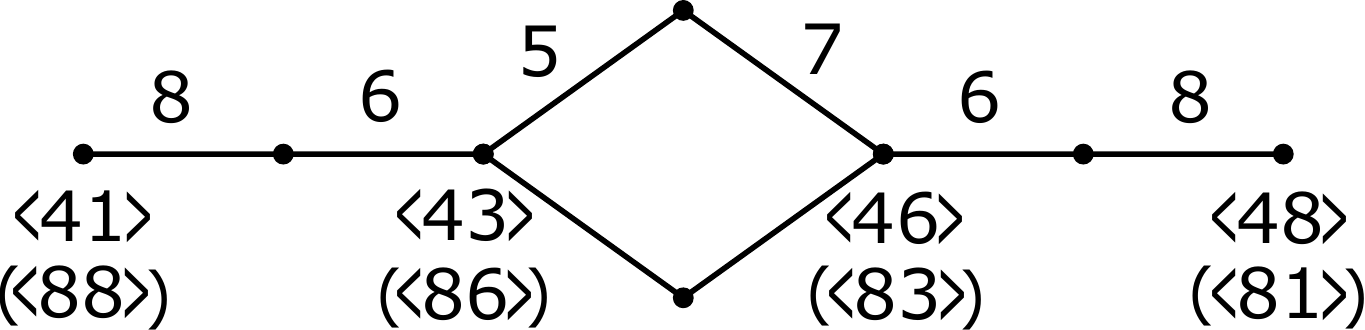}\vspace{1cm}\\
\mathrm{(a)} &~~~~~~& \mathrm{(b)}
\end{array}
$
\end{center}
 \caption{$N=8$ case. Diagrams near $\la 1 \ra$($\la 128 \ra$). Diagram (b) is the same as Figure \ref{fig:n4}.  }
 \label{fig:n8_decomp}
\end{figure}

\begin{figure}[ht!]
\begin{center}
$\begin{array}{ccc}
\includegraphics[width=7cm,clip]{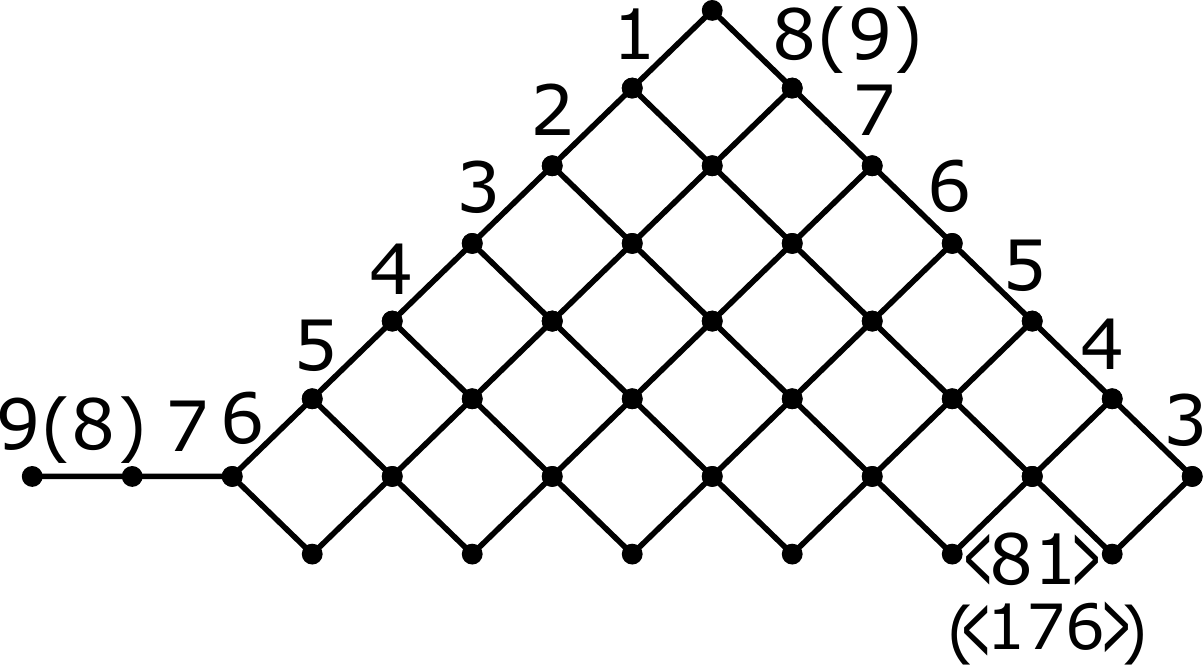}
&~~~~~~&
\includegraphics[width=7cm,clip]{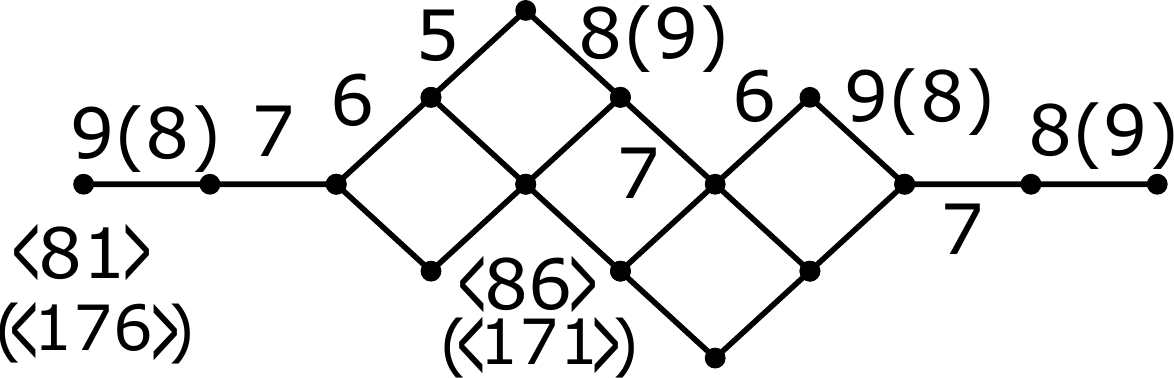}\vspace{1cm}\\
\mathrm{(a)} &~~~~~~& \mathrm{(b)}
\end{array}
$
\end{center}
 \caption{$N=9$ case. Diagrams near $\la 1 \ra$($\la 256 \ra$). Diagram (b) is the same as Figure \ref{fig:n5}.}
 \label{fig:n9_decomp}
\end{figure}

In Figure \ref{fig:n6} and Figure \ref{fig:n6v}, the vacuum which is connected to the maximum number of simple roots near $\la 1 \ra$ is $\la 11 \ra$. The vacuum structure near $\la 1 \ra$ decomposes into two diagrams as it is shown in Figure \ref{fig:n6_decomp}. From this we can conclude that $\la 11 \ra$ is connected to the maximum number of elementary walls. Figure (b) is the same as Figure \ref{fig:n2n3rts} (a).
In Figure \ref{fig:n6} and Figure \ref{fig:n6v}, the vacuum, which is connected to the maximum number of simple roots near $\la 32 \ra$ is $\la 22 \ra$. This can also be seen by decomposing the vacuum structure near $\la 32 \ra$. We get the same two-dimensional diagrams as
Figure \ref{fig:n6_decomp} by replacing $\la 11 \ra$ and $\la 12 \ra$ with $\la 22 \ra$ and $\la 21 \ra$. In this case the vacua on the left hand side are in the $x\rightarrow -\infty$ limit and the vacua on the right hand side are in the $x\rightarrow \infty$ limit. In the same manner, Figure \ref{fig:n7} and Figure \ref{fig:n7v} decompose as it is shown in Figure \ref{fig:n7_decomp}. The vacua which are connected to the maximum number of simple roots are $\la 22 \ra$ and $\la 43 \ra$. Figure \ref{fig:n7_decomp} (b) is the same as Figure \ref{fig:n2n3rts} (b).
$N=8$ and $N=9$ cases are depicted in Figure \ref{fig:n8_decomp} and in Figure \ref{fig:n9_decomp}. The vacuum structures repeat the four diagrams in Figure \ref{fig:n2n3rts}, Figure \ref{fig:n4} and Figure \ref{fig:n5}. 

\begin{figure}[ht!]
\begin{center}
$\begin{array}{ccc}
\includegraphics[width=7cm,clip]{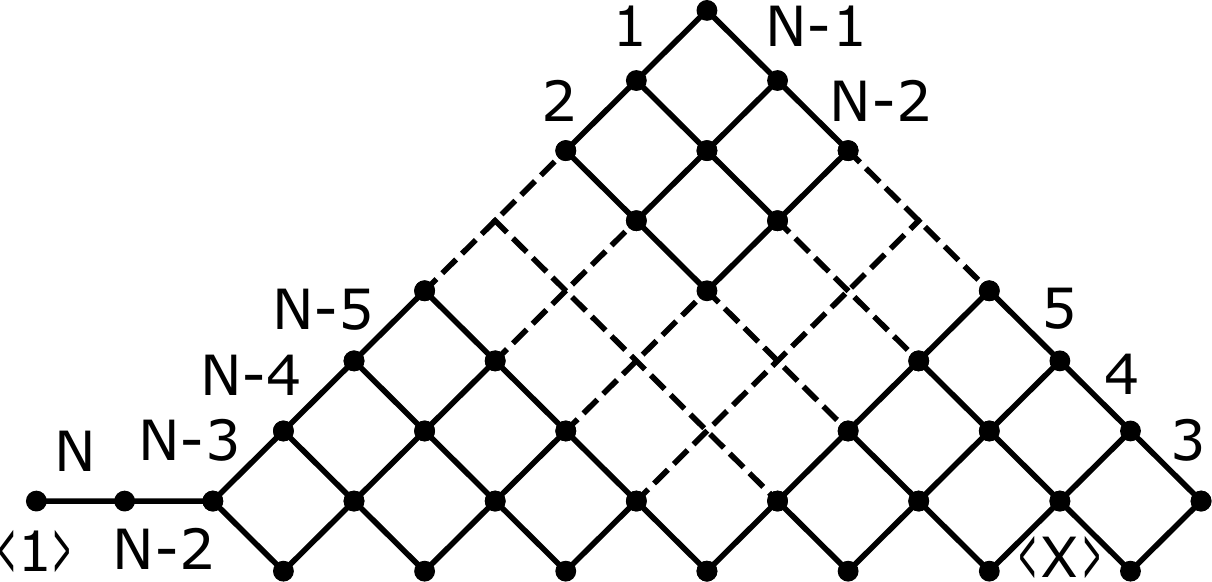}
&~~~~~~&
\includegraphics[width=7cm,clip]{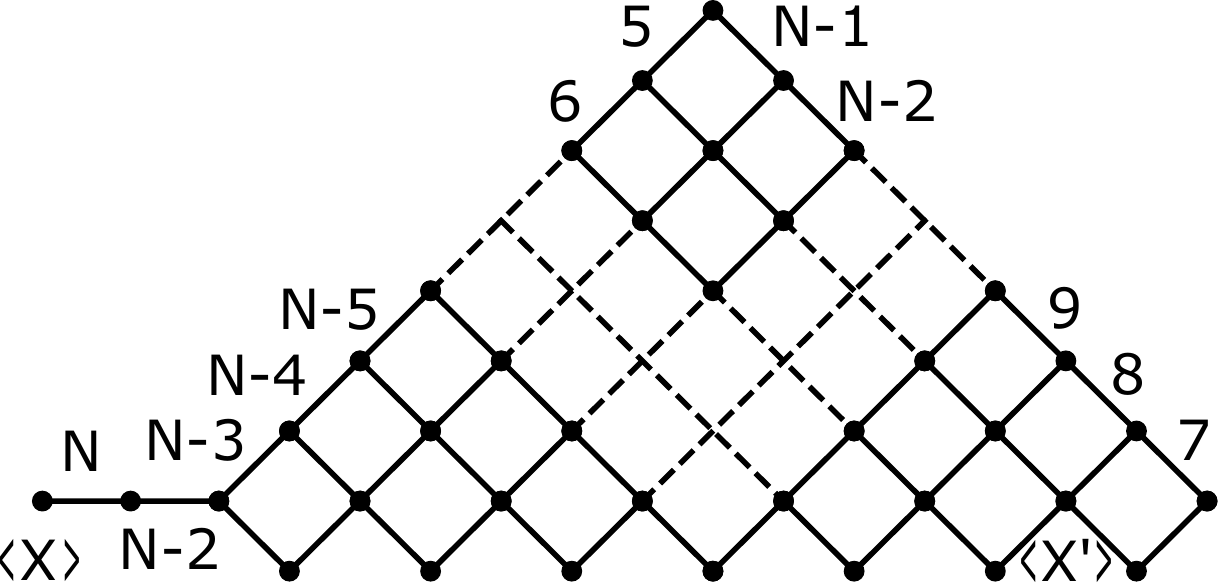}\vspace{1cm}\\
\mathrm{(a)} &~~~~~~& \mathrm{(b)} 
\end{array}
$
\end{center}
 \caption{First two diagrams of the vacuum structure near $\la 1 \ra$ for $N$.}
 \label{fig:n_decomp_a}
\end{figure}

\begin{figure}[ht!]
\begin{center}
$\begin{array}{ccc}
\includegraphics[width=7cm,clip]{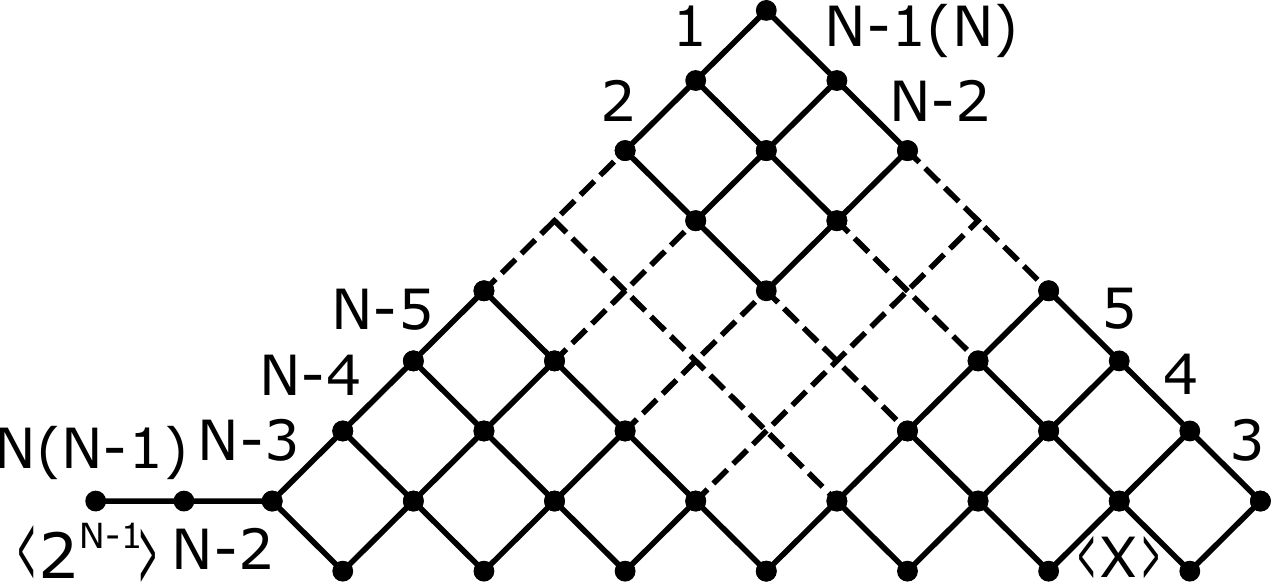}
&~~~~~~&
\includegraphics[width=7cm,clip]{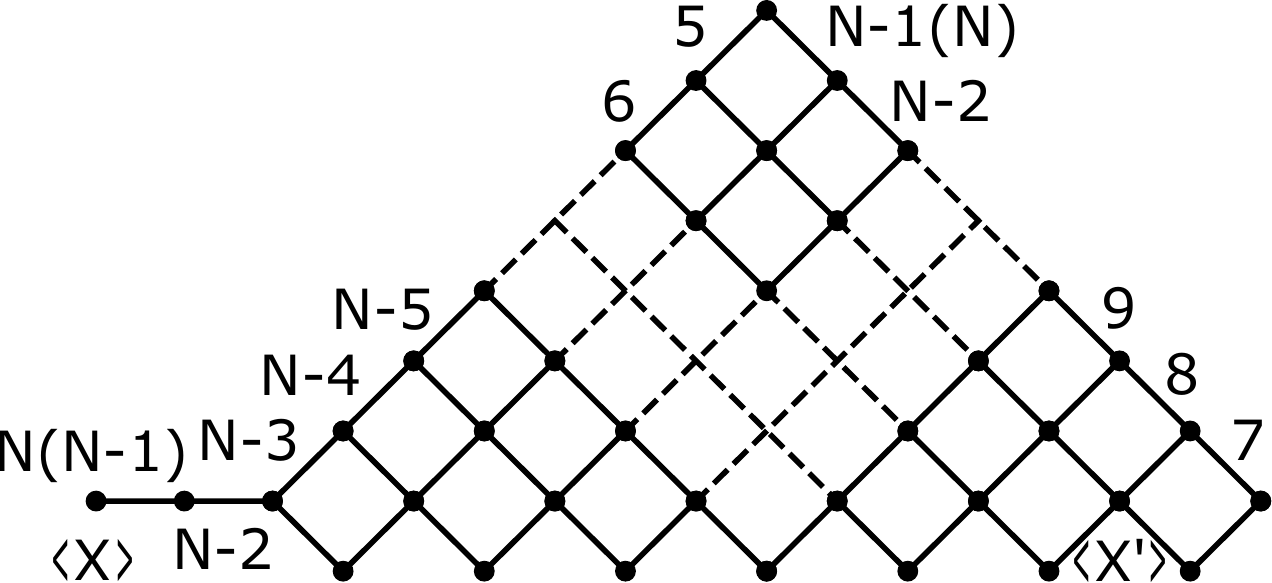}\vspace{1cm}\\
\mathrm{(a)} &~~~~~~& \mathrm{(b)} 
\end{array}
$
\end{center}
  \caption{First two diagrams of the vacuum structure near $\la 2^{N-1} \ra$ for $N$.}
 \label{fig:n_decomp_b}
\end{figure}

\begin{figure}[ht!]
\begin{center}
$\begin{array}{ccc}
\includegraphics[width=3cm,clip]{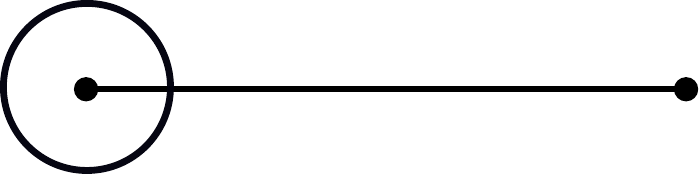}
&~~~~~~&
\includegraphics[width=6cm,clip]{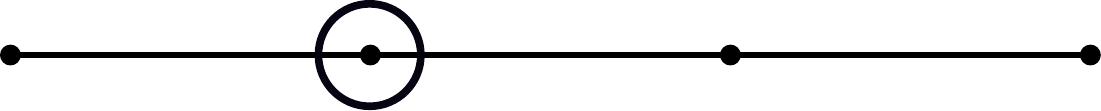}\vspace{1cm}\\
\mathrm{(a)} &~~~~~~& \mathrm{(b)}\\
\includegraphics[width=6cm,clip]{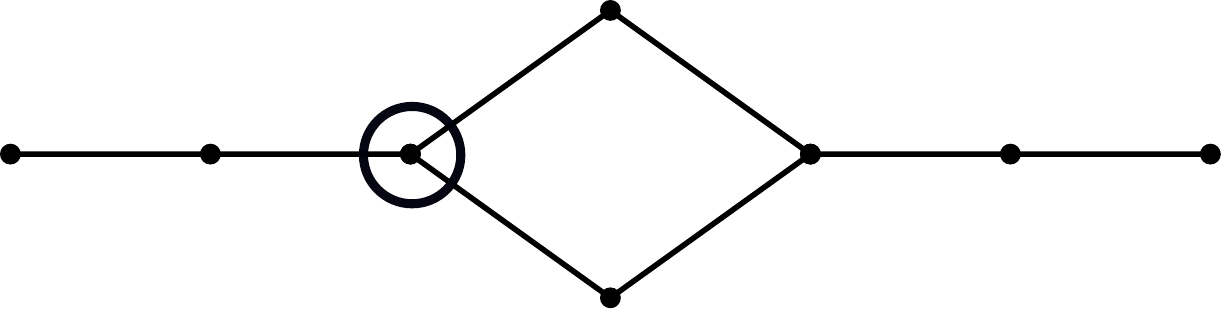}
&~~~~~~&
\includegraphics[width=7cm,clip]{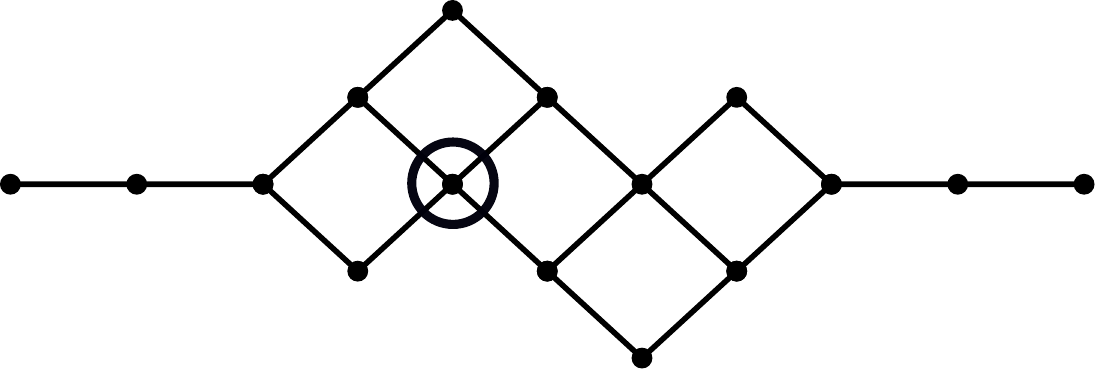}\vspace{1cm}\\
\mathrm{(c)} &~~~~~~& \mathrm{(d)} 
\end{array}
$
\end{center}
 \caption{Four types of vacuum structures. The circle indicates the vacuum which is connected to the maximum number of simple roots.  }
 \label{fig:4diagrams}
\end{figure}

The vacuum structure near $\la 1 \ra$ decomposes into the diagrams in Figure \ref{fig:n_decomp_a} while the vacuum structure near $\la 2^{N-1} \ra$ decomposes into the diagrams in Figure \ref{fig:n_decomp_b}. In both Figures, only the first two diagrams are shown. By repeating them all the cases fall into four categories. The vacuum which is connected to the maximum number of simple roots is circled in each diagram in Figure \ref{fig:4diagrams}.


\begin{figure}[ht!]
\begin{center}
\includegraphics[width=12cm,clip]{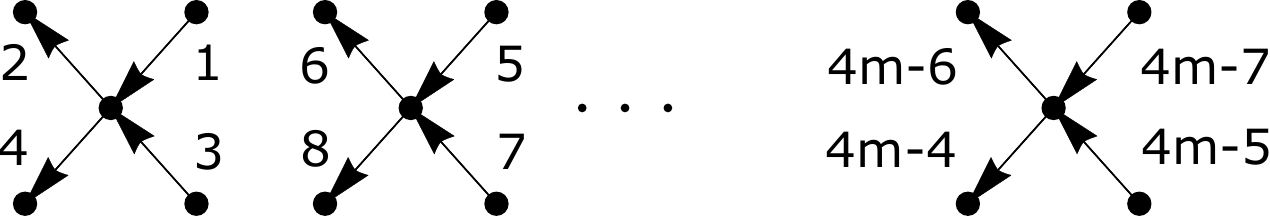}
\end{center}
 \caption{Common part of the vacuum structure near $\la A \ra$.}
 \label{fig:a1}
\end{figure}

\begin{figure}[ht!]
\begin{center}
\includegraphics[width=12cm,clip]{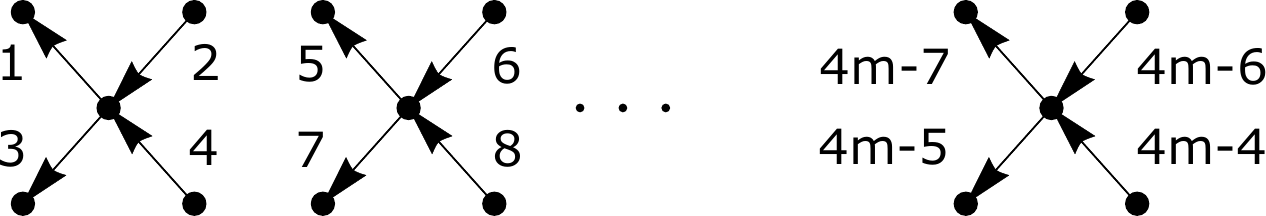}
\end{center}
 \caption{Common part of the vacuum structure near $\la B \ra$.}
 \label{fig:b1}
\end{figure}

\begin{figure}[ht!]
\begin{center}
$\begin{array}{ccc}
\includegraphics[width=5cm,clip]{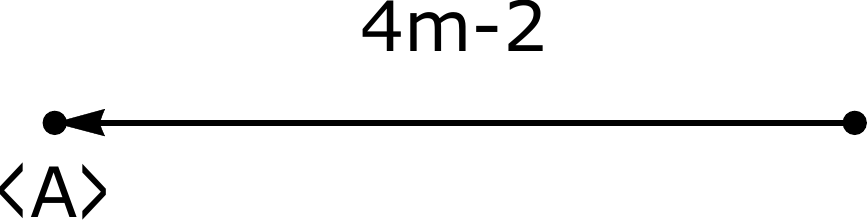}
&~~~~~~&
\includegraphics[width=5cm,clip]{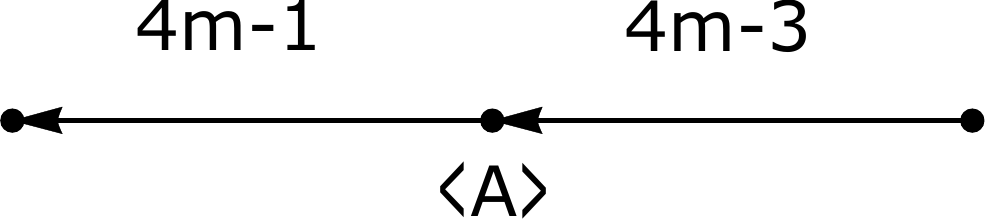}\\
~~ & ~~ & ~~ \\
\mathrm{(a)} &~~~~~~& \mathrm{(b)}  \\
~~ & ~~ & ~~ \\
~~ & ~~ & ~~ \\
\includegraphics[width=5cm,clip]{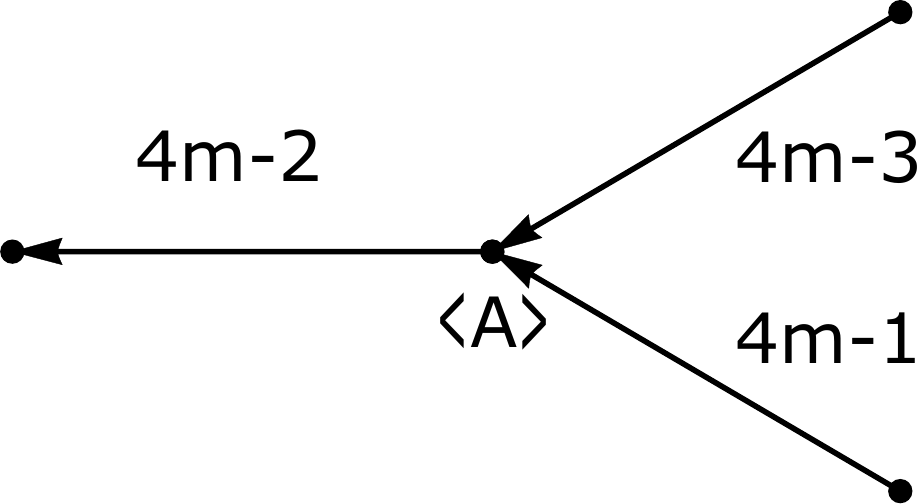}
&~~~~~~&
\includegraphics[width=5cm,clip]{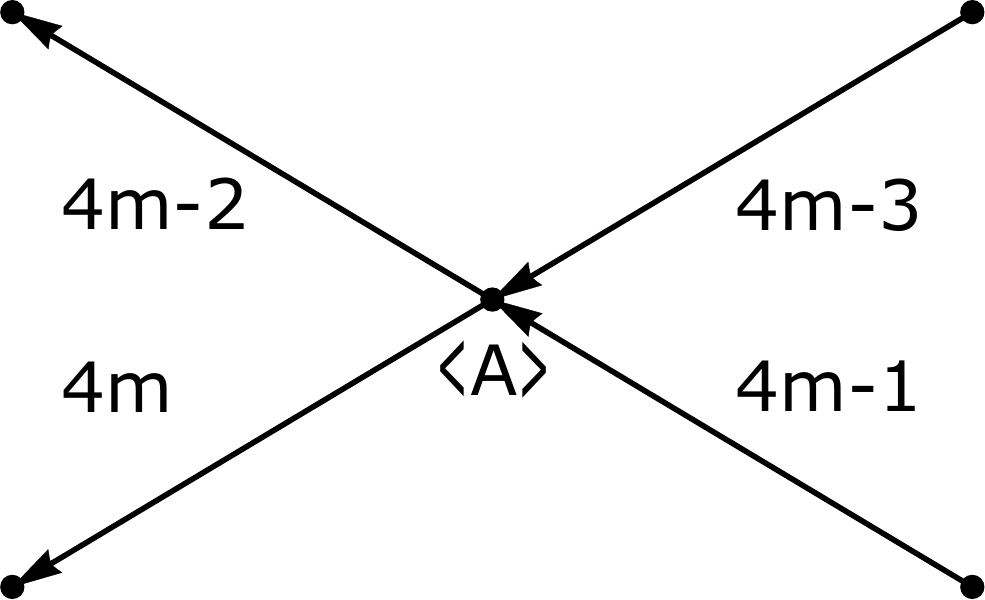}\\
~~ & ~~ & ~~ \\
\mathrm{(c)} &~~~~~~& \mathrm{(d)} 
\end{array}
$
\end{center}
 \caption{Rest part of the vacuum structure near $\la A \ra$.}
 \label{fig:a2}
\end{figure}

\begin{figure}[ht!]
\begin{center}
$\begin{array}{ccc}
\includegraphics[width=5cm,clip]{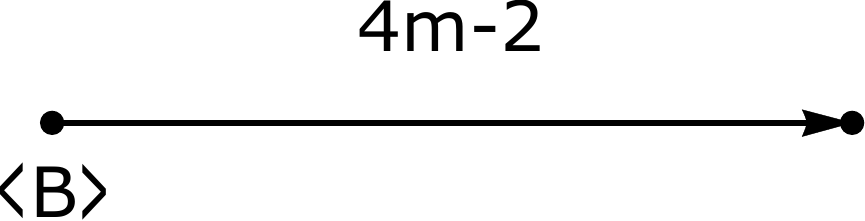}
&~~~~~~&
\includegraphics[width=5cm,clip]{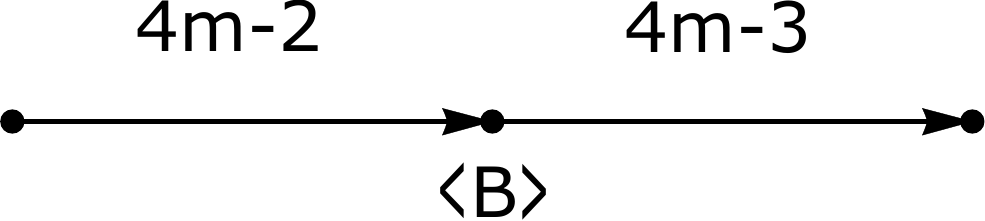}\\
~~ & ~~ & ~~ \\
\mathrm{(a)} &~~~~~~& \mathrm{(b)}  \\
~~ & ~~ & ~~ \\
~~ & ~~ & ~~ \\
\includegraphics[width=5cm,clip]{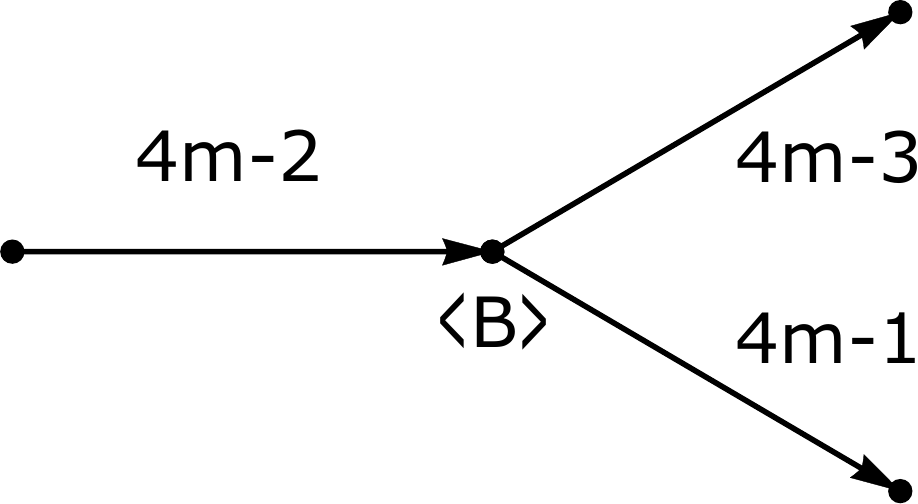}
&~~~~~~&
\includegraphics[width=5cm,clip]{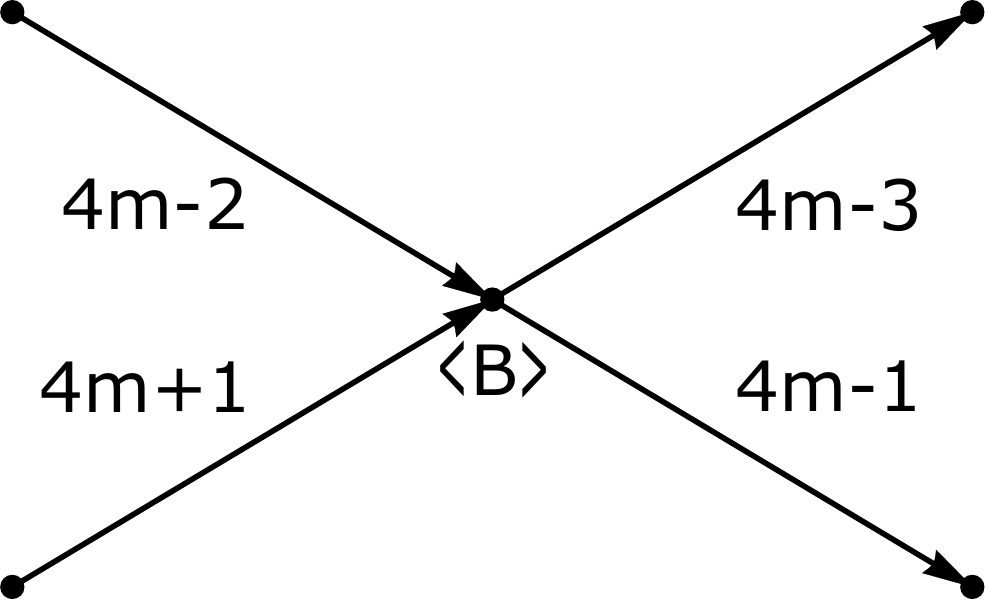}\\
~~ & ~~ & ~~ \\
\mathrm{(c)} &~~~~~~& \mathrm{(d)} 
\end{array}
$
\end{center}
 \caption{Rest part of the vacuum structure near $\la B \ra$.}
 \label{fig:b2}
\end{figure}

There are two vacua which are connected to the maximum number of elementary walls. $\la A \ra$ denotes the vacuum near $\la 1 \ra$ and $\la B \ra$ denotes the vacuum near $\la 2^{N-1} \ra$. The common parts of each vacuum structure is shown in Figure \ref{fig:a1} and Figure \ref{fig:b1}. The rest of the vacuum structure can be derived from Figure \ref{fig:4diagrams}. The vacuum structures near $\la A \ra$ and $\la B \ra$ are illustrated in Figure \ref{fig:a2} and Figure \ref{fig:b2} respectively.


\clearpage

Figure \ref{fig:a1} and Figure \ref{fig:a2} lead to the following:
\begin{itemize}
\item $N=4m-2$, $(m\geq 2)$
\bea
\Big\{ \underbrace{\vec{2},\vec{4},\cdots,\vec{4m-4}}_{2m-2} \Big\}
\leftarrow \la A \ra \leftarrow 
\Big\{\underbrace{\vec{1},\vec{3},\cdots,\vec{4m-5}}_{2m-2},\vec{4m-2}\Big\}  \label{eq:vacaa1}
\eea
\item $N=4m-1$, $(m\geq 2)$
\bea
\Big\{ \underbrace{\vec{2},\vec{4},\cdots,\vec{4m-4}}_{2m-2},\vec{4m-1} \Big\}
\leftarrow \la A \ra \leftarrow 
\Big\{\underbrace{\vec{1},\vec{3},\cdots,\vec{4m-5},\vec{4m-3}}_{2m-1}\Big\}  \label{eq:vacaa2}
\eea
\item $N=4m$, $(m\geq 2)$
\bea
\Big\{\underbrace{\vec{2},\vec{4},\cdots,\vec{4m-2} }_{2m-1}\Big\}
\leftarrow \la A \ra \leftarrow \Big\{\underbrace{\vec{1},\vec{3},\cdots,\vec{4m-1}}_{2m}\Big\} \label{eq:vacaa3}
\eea
\item $N=4m+1$, $(m\geq 2)$
\bea
\Big\{\underbrace{\vec{2},\vec{4},\cdots,\vec{4m} }_{2m}\Big\}
\leftarrow \la A \ra \leftarrow \Big\{\underbrace{\vec{1},\vec{3},\cdots,\vec{4m-1}}_{2m}\Big\}  \label{eq:vacaa4}
\eea
\end{itemize}
Each case with $m=2$ is shown in Figure \ref{fig:n6_decomp}, Figure \ref{fig:n7_decomp}, Figure \ref{fig:n8_decomp} and Figure \ref{fig:n9_decomp}. Let us assume that (\ref{eq:vacaa1}), (\ref{eq:vacaa2}), 
(\ref{eq:vacaa3}) and (\ref{eq:vacaa4}) are true. Then these are true for 
$m^\prime=m+1$ since it corresponds to adding one more diagram in Figure \ref{fig:a1}.
Therefore (\ref{eq:vacaa1}), (\ref{eq:vacaa2}), 
(\ref{eq:vacaa3}) and (\ref{eq:vacaa4}) are true.

Figure \ref{fig:b1} and Figure \ref{fig:b2} lead to the following:
\begin{itemize}
\item $N=4m-2$, $(m\geq 2)$
\bea
\Big\{\underbrace{\vec{1},\vec{3},\cdots,\vec{4m-5}}_{2m-2},\vec{4m-2}\Big\} 
\leftarrow \la B \ra \leftarrow 
\Big\{ \underbrace{\vec{2},\vec{4},\cdots,\vec{4m-4}}_{2m-2} \Big\}
 \label{eq:vacbb1}
\eea
\item $N=4m-1$, $(m\geq 2)$
\bea
\Big\{\underbrace{\vec{1},\vec{3},\cdots,\vec{4m-5},\vec{4m-3}}_{2m-1}\Big\} 
\leftarrow \la B \ra \leftarrow 
\Big\{ \underbrace{\vec{2},\vec{4},\cdots,\vec{4m-2}}_{2m-1} \Big\}
 \label{eq:vacbb2}
\eea
\item $N=4m$, $(m\geq 2)$
\bea
\Big\{\underbrace{\vec{1},\vec{3},\cdots,\vec{4m-1}}_{2m}\Big\} 
\leftarrow \la B \ra \leftarrow 
\Big\{ \underbrace{\vec{2},\vec{4},\cdots,\vec{4m-2}}_{2m-1} \Big\}
 \label{eq:vacbb3}
\eea
\item $N=4m+1$, $(m\geq 2)$
\bea
\Big\{\underbrace{\vec{1},\vec{3},\cdots,\vec{4m-1}}_{2m}\Big\} 
\leftarrow \la B \ra \leftarrow 
\Big\{ \underbrace{\vec{2},\vec{4},\cdots,\vec{4m-2}}_{2m-1},\vec{4m+1} \Big\}
 \label{eq:vacbb4}
\eea
\end{itemize}

Each case with $m=2$ is shown in Figure \ref{fig:n6_decomp}, Figure \ref{fig:n7_decomp}, Figure \ref{fig:n8_decomp} and Figure \ref{fig:n9_decomp}. Let us assume that (\ref{eq:vacbb1}), (\ref{eq:vacbb2}), 
(\ref{eq:vacbb3}) and (\ref{eq:vacbb4}) are true. Then these are true for 
$m^\prime=m+1$ since it corresponds to adding one more diagram in Figure \ref{fig:b1}.
Therefore (\ref{eq:vacbb1}), (\ref{eq:vacbb2}), 
(\ref{eq:vacbb3}) and (\ref{eq:vacbb4}) are true.

The vacuum structure is
\bea
\vec{N}\leftarrow \cdots \leftarrow \la A \ra \leftarrow \cdots \leftarrow \la B \ra \leftarrow \cdots
\leftarrow \vec{N}
\eea
for even $N$ and
\bea
\vec{N}\leftarrow \cdots \leftarrow \la A \ra \leftarrow \cdots \leftarrow \la B \ra
\leftarrow \cdots \leftarrow \vec{N-1}
\eea
for odd $N$.

Therefore (\ref{eq:4m-2}), (\ref{eq:4m-1}), (\ref{eq:4m}) and (\ref{eq:4m+1}) are proved.

\end{document}